\documentclass[iop]{emulateapj} 

\usepackage{graphics} 
\usepackage{amsmath}

\newcommand{\um}{${\rm \mu m}$~}
\newcommand{\mm}{${\rm \mu m}$}
\newcommand{\epseri} {$\epsilon$ Eri~}

\long\def\symbolfootnote[#1]#2{\begingroup%
\def\thefootnote{\fnsymbol{footnote}}\footnote[#1]{#2}\endgroup}

\slugcomment{accepted by AJ}

\shorttitle{35 \um image of \epseri}
\shortauthors{Su et al.}

\begin{document}

\title{The Inner 25 AU Debris Distribution in the $\epsilon$ Eri System}

\author{Kate Y. L. Su\altaffilmark{1}, James M. De Buizer\altaffilmark{2}, George H. Rieke\altaffilmark{1}, 
Alexander V. Krivov\altaffilmark{3}, Torsten L\"ohne\altaffilmark{3}, Massimo Marengo\altaffilmark{4}, Karl R. Stapelfeldt\altaffilmark{5}, Nicholas P. Ballering\altaffilmark{1}, and William D. Vacca\altaffilmark{2} }

\affil{$^1$ Steward Observatory, University of Arizona, 933 N Cherry Ave., Tucson, AZ 85721, USA}
\affil{$^2$ SOFIA-USRA, NASA Ames Research Center, MS 232-12, Moffett Field, CA 94035, USA}
\affil{$^3$ Astrophysikalisches Institut und Universit\"atssternwarte, Friedrich-Schiller-Universit\"at Jena, Schillerg\"a{\ss}chen~2--3, 07745 Jena, Germany}              
\affil{$^4$ Department of Physics \& Astronomy, Iowa State University, Ames, IA 50011, USA}
\affil{$^5$ Jet Propulsion Laboratory, California Institute of Technology, 4800 Oak Grove Drive, Pasadena, CA 91109, USA}

\begin{abstract}

Debris disk morphology is wavelength dependent due to the wide range
of particle sizes and size-dependent dynamics influenced by various
forces. Resolved images of nearby debris disks reveal complex disk
structures that are difficult to distinguish from their spectral
energy distributions.  Therefore, multi-wavelength resolved images of
nearby debris systems provide an essential foundation to understand
the intricate interplay between collisional, gravitational, and
radiative forces that govern debris disk structures. We present the
{\it SOFIA} 35 \um resolved disk image of $\epsilon$ Eri, the closest
debris disk around a star similar to the early Sun. Combining with the
{\it Spitzer} resolved image at 24 \um and 15--38 \um excess spectrum,
we examine two proposed origins of the inner debris in $\epsilon$ Eri:
(1) in-situ planetesimal belt(s) and (2) dragged-in grains from the
cold outer belt. We find that the presence of in-situ dust-producing
planetesmial belt(s) is the most likely source of the excess emission
in the inner 25 au region. Although a small amount of dragged-in
grains from the cold belt could contribute to the excess emission in
the inner region, the resolution of the {\it SOFIA} data is high
enough to rule out the possibility that the entire inner warm excess
results from dragged-in grains, but not enough to distinguish one
broad inner disk from two narrow belts.

\end{abstract}


\section{Introduction}

Debris disks are integral parts of planetary systems. They are
produced when larger objects, e.g. planets, stir planetesimal belts,
causing a cascade of collisions that break minor bodies down into
dust. More than 400 debris disks are known, providing a rich resource
to study planetary system evolution and architecture
\citep{matthews14}. However, the majority only have photometric points
defining a general spectral energy distribution (SED). SEDs measure
temperature, but grains with different optical properties can have the
same temperature at different distances from a star, making SED
modeling degenerate. Resolved images are essential to eliminate this
degeneracy. The few systems that are close enough to be well resolved
provide the foundation for the entire effort to interpret debris disk
behavior in terms of the underlying planetary configuration. Because
the huge range of particle sizes (from sub-$\mu$m to mm/cm sizes)
produced in debris disks results in size-dependent dynamics influenced
by various forces (radiation and drag), the observed disk structures
are wavelength-dependent (e.g., \citealt{wyatt06}). Therefore,
multi-wavelength observations are essential to understand the
intricate interplay governing debris disk structures.

The two benchmark nearby debris disks are not around sun-like stars,
but are around the early A-stars Fomalhaut and Vega. These are aptly
termed the debris disk twins, not only because of the similar stellar
types, but their similar ages ($\sim$450 Myr), the evidence for warm
belts, and their prominent cold belts \citep{su13,su16}. They have a
large gap between their warm and cold dust belts, a possible signpost
for multiple, low-mass planets beyond the water-ice lines that
typically lie near the warm belts (e.g., \citealt{quillen06};
\citealt{su13}). The high temperatures ($\sim$9,000 K) and
luminosities ($\sim$16 and $\sim$30 L$_{\sun}$ respectively) of these
stars subject their debris dust to different environments than for
dust around the Sun -- different not only in the radially-dependent
equilibrium temperature, but also in the roles of photon pressure,
magnetic fields, and stellar winds. Given that planetesimal belts
probably form near the primordial ice line, the relatively weak
dependence of this location on pre-main-sequence stellar luminosity
\citep{kennedy08} also potentially contributes to significant
differences in the planetesimal belt environments.

Therefore, it is important to contrast debris systems around stars
more like the Sun with those around Fomalhaut and Vega. Within 5 pc,
$\tau$ Ceti (3.65 pc, \citealt{vanleeuwen07}) and \epseri 
(3.22 pc, \citealt{vanleeuwen07}) are the only two low-mass stars
with prominent debris disks. The age of $\tau$ Ceti (5.8 Gyr) results
in a faint disk \citep{sierchio14}, greatly limiting the detectability
of detailed disk structures. \epseri provides a better translation
from the debris properties of Vega and Fomalhaut to the environment of
the solar system. It is at a similar age (400--800 Myr,
\citealt{difolco04,mamajek08}) as Vega and Fomalhaut, but its
temperature, mass, and luminosity (5,100 K, $M_{\ast}$=0.82 M$_\odot$,
and $L_{\ast}$= 0.34 L$_\odot$) suggest that $\epsilon$ Eri should
have similar properties with regard to magnetic field, stellar winds,
and UV output as the early Sun.

Although the \epseri debris disk has been resolved at multiple
wavelengths, the structure of its debris system remains
controversial. At 850 $\mu$m, JCMT/SCUBA revealed a nearly face-on,
clumpy Kuiper-belt-like ring at a radius of $\sim$64 au
\citep{greaves98}. The clumpy structure has been interpreted as
evidence for an unseen planet interior of the cold ring
\citep{ozernoy00,quillen02,deller05}, but the perturbing planet
remains undetected (e.g., \citealt{janson15} and references
therein). The large cold ring at 64 au is confirmed in the mm
\citep{lestrade15,macgregor15}. From these images, it appears that
many of the clumps can be ascribed to chance alignment with background
galaxies \citep{chavez16}.

{\it Spitzer }imaging and spectroscopic data combined with a SED model
suggest the existence of two distinct belts in its inner 25 au region
\citep{backman09}.  In the Backman model, the inner warm belt is
similar in location to our own Asteroid belt located at $\sim$3 au,
while the outer warm belt lies close to where Uranus orbits in our
solar system ($\sim$20 au). The exact location and the width of the
two inner warm belts as proposed by \citet{backman09} are only
constrained by marginally resolved images and the SED modeling with
assumed grain properties, and could be uncertain by factors of two.
Recent {\it Herschel } far-infrared images of the system suggest the
outer warm belt may be as close as 12--16 au \citep{greaves14}.

\citet{hatzes00} reported the detection of a planet, \epseri b, whose
orbit \citep{benedict06} may cross the innermost warm belt proposed by
\citet{backman09}, leading to an unstable configuration
\citep{brogi09}.  To avoid this difficulty, \citet{reidemeister11}
instead suggested that the warm excess originates from small
($\lesssim$10 $\mu$m) grains in the cold outer belt, which are
transported inward by Poynting-Robertson (P-R) and stellar wind
drag. According to this hypothesis, the disk surface density is
expected to be relatively flat between the warm and cold components
while the radially dependent dust temperatures result in a centrally
peaked 24 $\mu$m image. Under this model, there is no need for an
inner planetesimal belt as the source of the warm dust.  However,
\citet{butler06} suggest that the planet's orbit is much less
eccentric, even consistent with being circular, which might make this
model unnecessary. There is also controversy over whether the planet
is real \citep{zechmeister13}, suggesting an alternative solution to
the dilemma.

To better understand the debris distribution in the inner 25 au region
of $\epsilon$ Eri, we obtained {\it SOFIA }35 $\mu$m images of this
system with a resolution of 3$\farcs$4. The details of the
observations and data reduction, including the archival observations
of calibrators, are presented in Section 2. A detailed
characterization of the {\it SOFIA }35 $\mu$m Point Spread Function
(PSF) allows us to assess the disk extent at this wavelength, and show
that the emission is centrally peaked but extends beyond two
resolution elements, and then drops off quickly outside 10\arcsec\
(Section 3). We analyze the disk radial profiles in the mid-infrared
(with additional archival {\it Spitzer } MIPS 24 $\mu$m and IRS data)
in Section 4. In Section 5, we use the 24 and 35 $\mu$m disk radial
profiles to test three different debris distributions proposed in the
literature, and suggest that the inner warm dust originates from one
or two planetesimal belts lying within 25 au of $\epsilon$ Eri. We
discuss the degeneracy in our choices of model parameters in Section 6
and conclude the paper in Section 7.

\section{FORCAST Observations and Data Reduction}

\epseri was observed with the NASA Stratospheric Observatory for
Infrared Astronomy ({\it SOFIA}, \citealt{gehrz09,young12}) during cycle 2
and 3 using the FORCAST instrument \citep{herter12} in the F348 filter
($\lambda_{eff}$= 34.8 \mm, $\Delta\lambda$= 3.8 \mm) of the Long Wave
Camera (LWC), resulting in a 3\farcm4$\times$3\farcm2 instantaneous field
of view with 0\farcs768 pixels after distortion correction. The
chopping was done with the Nod-Match-Chop (NMC) configuration with a chop
throw of 60\arcsec\ and a chop angle of 30\arcdeg\ in the array
coordinates to cancel atmospheric emission. A five-point dither
pattern with an offset of 10\arcsec\ in both RA and Dec directions was
used to correct for array artifacts. Details about the observations
are given in Table \ref{tab1_obs}.  The data were calibrated and
reduced with the pipeline software (ver.~1.1.0) by the {\it SOFIA} Science
Center.

To assess the presence of any extended emission structure around
$\epsilon$ Eri, we also performed similar data reduction on archival
calibration data obtained with the F348 filter during cycle 2 and 3,
which include a handful of stellar calibrators (blue PSF sources) and
the asteroid Ceres (red source).

\begin{deluxetable*}{cccccccc}
\tablewidth{0pc}
\footnotesize 
\tablecaption{Observational Log \label{tab1_obs}}
\tablehead{
\colhead{\#} & \colhead{Flight} & \colhead{Date} & \colhead{UT Time}& \colhead{FWHMx} & \colhead{FWHMy} & \colhead{$\sigma_{sky}$} & \colhead{Integration$^{\dagger}$} \\
\colhead{}   & \colhead{}  & \colhead{}   & \colhead{}   & \colhead{[arcsec]} & \colhead{[arcsec]}  & \colhead{[mJy arcsec$^{-2}$]}& \colhead{[sec]} 
}
\startdata 
 1& 190	& 2015-01-29	&  05:02:38.8	&  4.51	&  3.15	&   6.64	&  1162	\\   
 2& 190	& 2015-01-29	&  05:29:09.6	&  4.32	&  3.48	&   5.75	&  1290	\\   
 3& 190	& 2015-01-29	&  05:59:43.5	&  4.26	&  3.43	&   6.37	&  1162	\\   
 4& 190	& 2015-01-29	&  06:25:51.9	&  3.53	&  3.11	&   7.10	&  1032	\\   
 5& 190	& 2015-01-29	&  06:52:34.9	&  3.64	&  3.28	&   6.21	&  1162	\\   
 6& 190	& 2015-01-29	&  07:19:24.5	&  4.60	&  2.64	&   7.26	&  774	\\   
 7& 191	& 2015-02-04	&  04:08:12.2	&  4.42	&  3.77	&   6.58	&  1678	\\   
 8& 191	& 2015-02-04	&  04:36:37.4	&  3.57	&  3.26	&   7.11	&  1420	\\   
 9& 254	& 2015-11-04	&  06:44:39.5	&  3.79	&  3.57	&   2.69	&  4368	\\   
10& 258	& 2015-11-13	&  06:32:55.6	&  4.70	&  3.91	&   3.12	&  4048	  
\enddata
\tablenotetext{$^\dagger$}{These are the final combined, on-source integration time. }
\end{deluxetable*}

\subsection{$\epsilon$ Eri}

The pipeline-produced Level-3 data products (i.e., nod-subtracted,
dithers aligned, flux calibrated merged data) were the basis for
further analysis (coadding and custom background subtraction). The
\epseri observations consist of 10 Level-3 images. Visual inspection
of these images found one of the Level-3 products (\#6, with the
shortest total on-source integration time) has elongated image shapes.
We did not use these data for the final coadd. We coadded the good
images at subpixel levels with two different registration methods.
The first method was to define the centroid of the source by fitting a
2-D Gaussian profile\footnote{Note that the FORCAST point spread
function is better described by a Moffat function. Here we use a
Gaussian function as an approximation; the centroid of the source
should not be affected.} to the 10$\times$10 pixel area centered on
the source. The measured Full Width at Half Maximum (FWHM) of the
source in each of the images is given in Table 1. The average FWHM is
4\farcs1$\times$3\farcs4 $\pm$ 0\farcs4$\times$0\farcs3; on average,
there is 9\% variation in source FWHM among the 9 good images. The
second method was to determine the sub-pixel shifts by cross
correlating the central 10$\times$10 pixel area centered around the
source. In both methods, each of the images was registered in
sub-pixel levels, and then coadded with weights determined by its
integration time.

The flatness of the background in the vicinity of the target is an 
important factor to assess the extension of the \epseri disk. We used a
custom sky subtraction program to take out the large-scale background
structure in the final coadded data by fitting a low-power,
two-dimensional polynomial on the coadded image with the source region
(central 38\farcs4 region) masked out.  The final sky-subtracted
coadded data are slightly different depending on the registration
methods. We will discuss the subtle difference in Section 3.2.

\subsection{PSF Calibrators - Ceres and Stellar Sources}

The {\it SOFIA } PSF is not as stable as space-based observatories
given that it is an airborne facility. Therefore, it is important to
evaluate the instrument PSF using point-source observations like Ceres
and stellar calibrators. Ceres was observed eight times with the F348
filter as a low-temperature flux calibrator during FORCAST flights in
2015. In addition, stellar calibration observations with the same
filter in 2015 are also included in our analysis: six $\alpha$ Boo,
two $\alpha$ Tau, two $\beta$ UMi, and one $\gamma$ Dra. We determined
the PSF variation by comparing the FWHM of these data using the same
method as in $\epsilon$ Eri. The average FWHM of the Ceres data is
3\farcs61$\times$3\farcs42 ($\pm$0\farcs28) with a variation of
$\sim$8\%.  The average FWHM of the stellar calibrator data is
3\farcs37$\times$3\farcs15 ($\pm$0\farcs20) with a variation of
6\%. This suggests that the typical PSF variation in the SOFIA data is
6--8\% for the central core (bright) region. We then combined the
stellar observations to build a high signal-to-noise (S/N) PSF to
assess the variation in the wing (faint) part of the PSF. Since the
stars have various brightnesses, and these data were obtained at
various altitudes (not flux calibrated), each individual Level-3
mosaic was first normalized before combining. The normalization is
based on the core flux within a small aperture (radius of 2.5 pixels =
1\farcs92).  Since the S/N in each individual image is relatively high
for these bright targets, the centroiding methods make no difference
in the final coadded data. We generated two final coadded PSFs: one
for Ceres and one for all stellar calibrators together. We also
applied an additional sky subtraction as in the \epseri data for both
coadded PSF data sets.  Since both PSFs and our \epseri data are
coadded from many individual observations, the variation in the PSF
should average out. Based on the final combined PSF data, it appears
that Ceres is slightly broader than the stellar calibrator (as judged
by the measured FWHM). A detailed comparison between the stellar and
Ceres PSFs is given in Section 3.1.

\section{FORCAST Mid-Infrared Imaging Results}

\subsection{FORCAST 35 \um Point-Spread Function}

\begin{figure}[b!] 
  \figurenum{1}
  \epsscale{1.0}
  \label{psfcomp} 
  \plotone{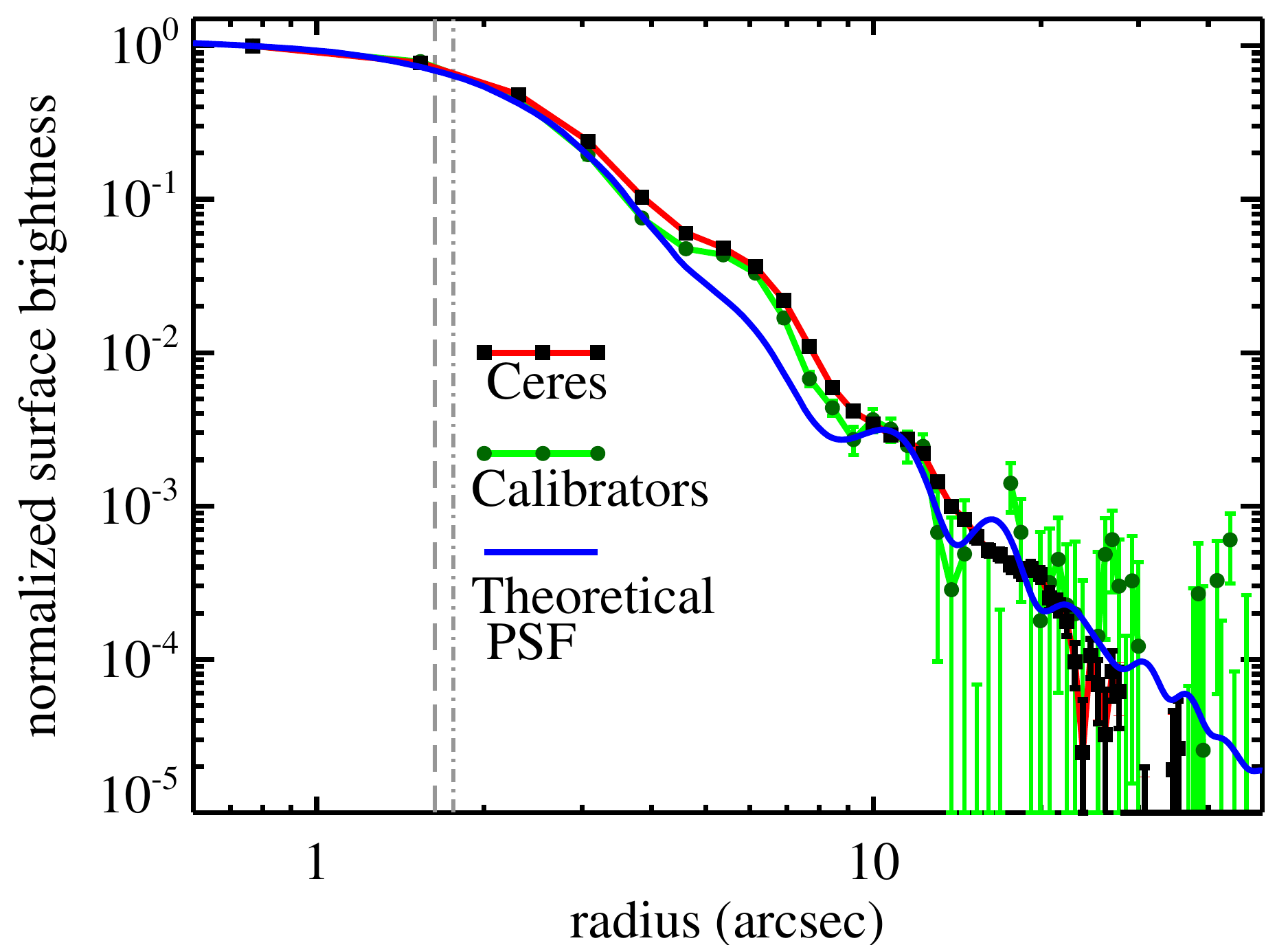}
  \caption{Radial profiles of the point spread function (PSF) for
FORCAST/SOFIA at 34.8 \um obtained with the Ceres and stellar
calibrators. The theoretical PSF profile (blue solid line) computed by
our custom IDL code is also shown for comparison. The vertical dashed
and dotted-dash lines mark the FWHM of the stellar and Ceres PSFs,
respectively. Our hybrid PSF is a combination of the stellar
calibrators (for the region inside 10\arcsec) and the theoretical PSF
(for the region outside 10\arcsec) (details see Sec 3.1). }
\end{figure}

To evaluate the FORCAST 35 \um PSF, we generated a theoretical PSF
using a custom IDL code with inputs of a wavelength and jitter value
to mimic the actual observations.  We generated 50 PSFs at wavelengths
across the F348 filter, and coadded them with weightings according to
the filter transmission to get a composite PSF. We found that a jitter
value of 1\farcs75 and a boxcar smooth factor of 3.4 pixels produce a
good match in terms of measured FWHM with the observed stellar PSF.

\begin{figure*}
  \figurenum{2}
  \label{coadd_comp} 
  \plottwo{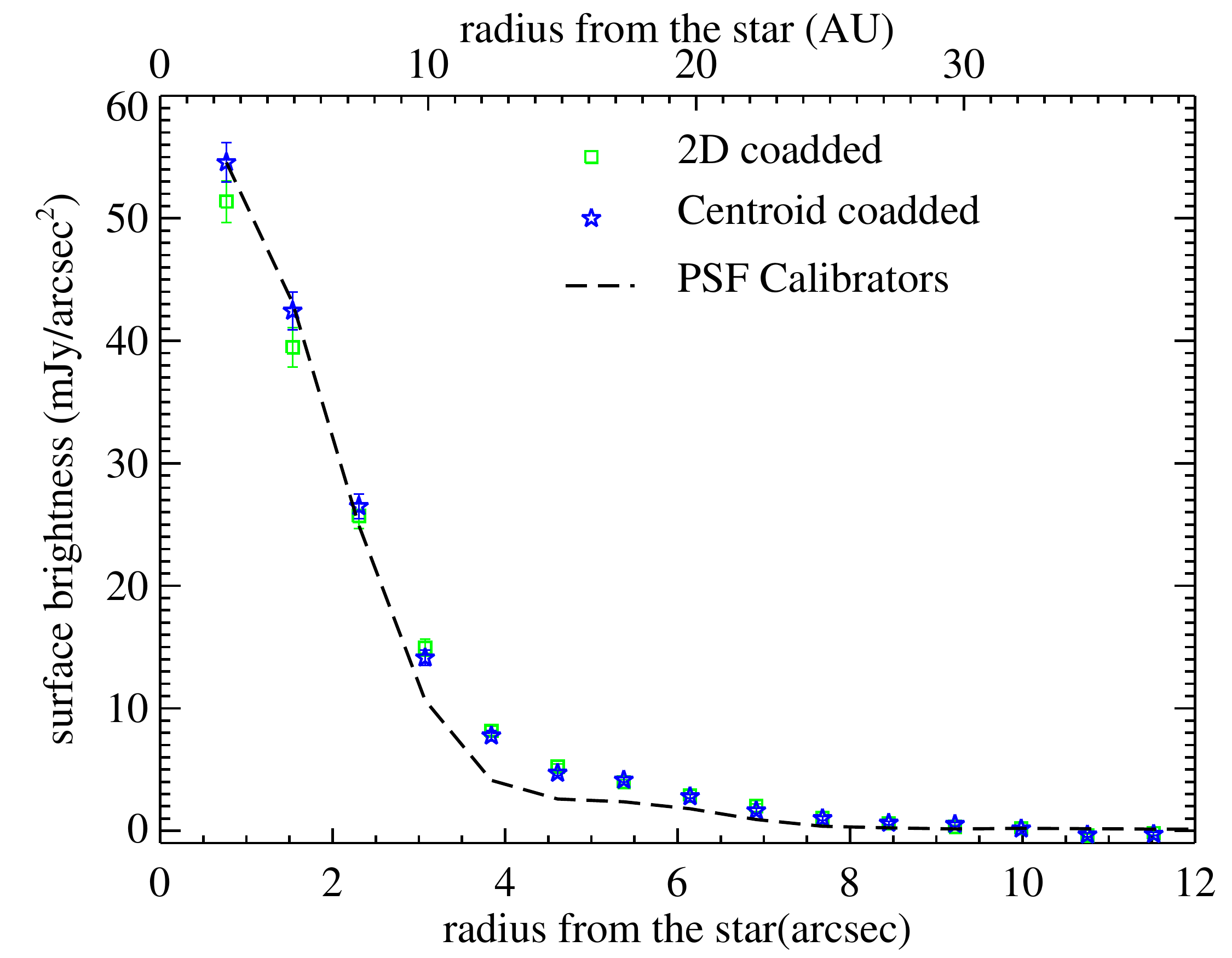}{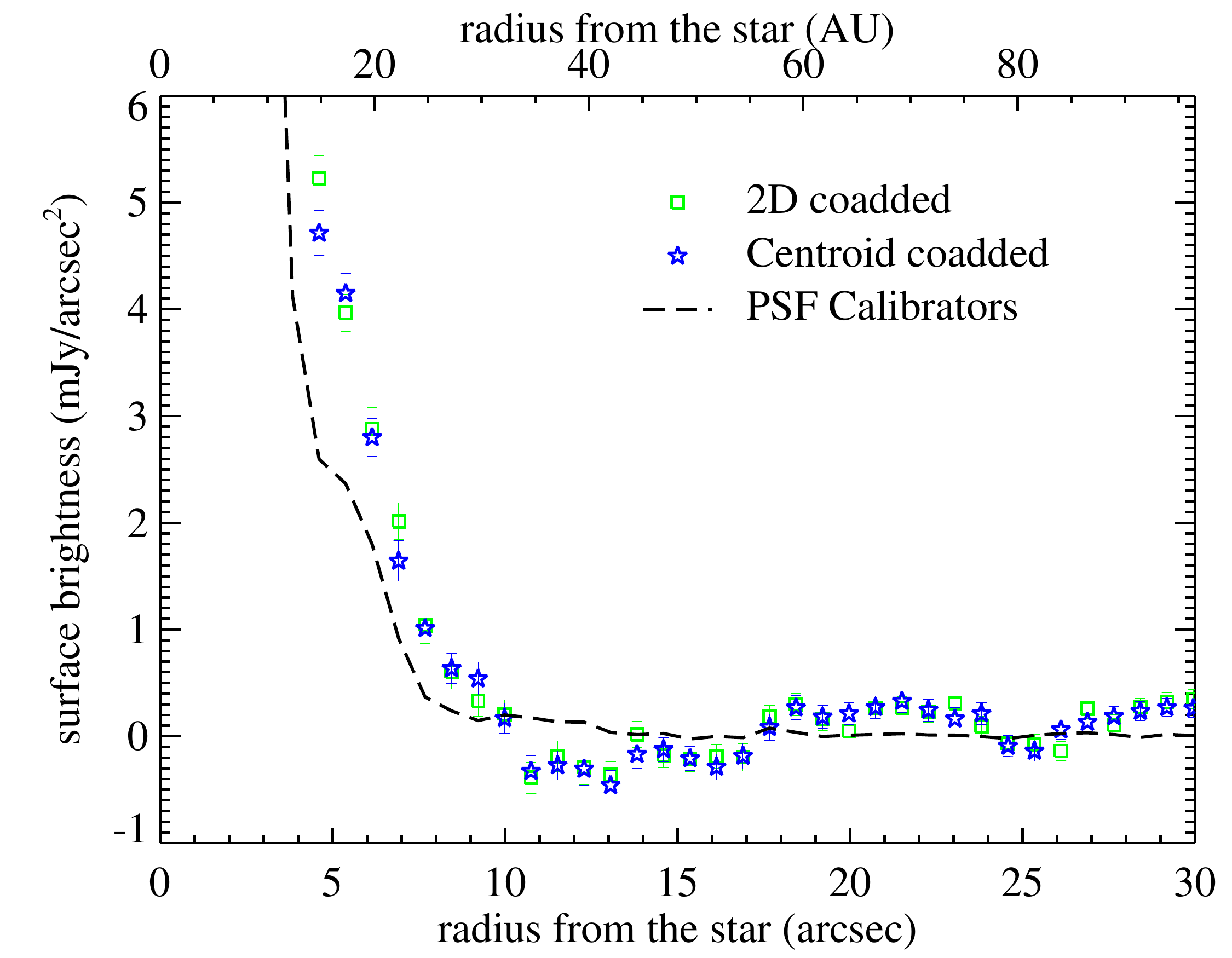}
  \caption{The comparison of the radial profiles for $\epsilon$ Eri
using two different registration methods. The hybrid PSF profile
(normalized to the peak, which is 21\% higher than the expected
photosphere) is shown as the black dashed line. The left panel covers the
central 12\arcsec\ region, while the right panel shows the full radial
range up to 30\arcsec\ but with a smaller vertical range of flux. All
error bars are shown as 1$\sigma$. There are some coherent up and down
patterns outside 10\arcsec, which are consistent with the flat
fielding residuals and within 3$\sigma$ of being zero. }
\end{figure*}

We compared the two observed PSFs and the theoretical PSF in terms of
azimuthally averaged radial profiles computed as follows. We first
created a series of concentric rings with a width of 1 pixel
(0\farcs768) about the source centroid (determined by the 2-D Gaussian
fit), and computed the average value of all the pixels that fall in
each ring. The measurement error at each radius is the standard
deviation of all pixels in that ring, divided by the square root of
the number of pixels in the ring. Since the FORCAST data are in the
background limited regime, the background noise also contributes to
the average flux measurement error. The background noise per pixel is
found by computing the standard deviation of the pixels on a part of
the blank region away from the source.  The background noise per ring
(i.e. background noise error) is estimated by taking the background
noise per pixel and dividing it by the square root of the number of
pixels in the ring. The total error in the average flux measurement
per ring, therefore, is the measurement error and the background noise
error added in quadrature. Figure \ref{psfcomp} shows the normalized
radial profiles for the PSF characterization. The high S/N of the
Ceres data enable us to track the radial profile up to 25\arcsec\ from
the center, achieving a dynamical range of 10$^{4}$. Overall, the
observed PSFs (both Ceres and calibrators) match the theoretical one
very well, except for the region at radii of 4\arcsec--10\arcsec.
The exact reason of the mismatch is unknown, but probably related
to how the data were taken and combined. As shown in Section 2.2, the
Ceres PSF is slightly broader than the stellar PSF by 8\% in the
measured FWHM.  Being a red source, Ceres is expected to be slightly
larger due to diffraction across the bandwidth of the filter. However,
the color difference can only account for a 2\% difference between
5000 K and 200 K sources. Ceres had an apparent diameter of 0\farcs7
around the time of observations, which can account for an additional
2\% difference in measured FWHM. The rest of this discrepancy is most
likely due to telescope tracking errors unique to the Ceres
observations. The {\it SOFIA} telescope uses a different technique to
track on non-sidereal targets and the tracking can be slightly less
accurate than sidereal tracking. Despite its high S/N, for this reason
we cannot simply use Ceres as our PSF standard for model convolution
(see Section 5). Instead, we constructed a hybrid PSF with the core
(inner 10\arcsec) high S/N) region using the observed stellar PSF and
the wing (outer 10\arcsec) region using the (noiseless) theoretical
PSF. We used this hybrid PSF for all of the following analysis.

\subsection{FORCAST 35 \um Image of $\epsilon$ Eri}

The subtle difference between the two registration methods in
combining the \epseri data is best shown in the measured FWHM and the
azimuthally averaged radial profiles for the image of the star (Figure
\ref{coadd_comp}). The left panel of Figure 2 shows the central
12\arcsec\ region, and the right panel shows the full range up to
30\arcsec. The centroiding method gives a slightly sharper image by
$\sim$3.5\% (FWHM = 3\farcs63$\times$3\farcs48), but the surface
brightness agrees within 1$\sigma$ as shown in the radial profiles.
The profile outside 10\arcsec\ is within 3$\sigma$ of zero and is
consistent with no signal within the expected uncertainties in flat
fielding. Therefore, we only concentrate our further analysis for the 
region inside a radius of 10\arcsec\ around $\epsilon$ Eri.

\begin{figure*}
  \figurenum{3}
  \label{epsEri_forcast} 
  \plotone{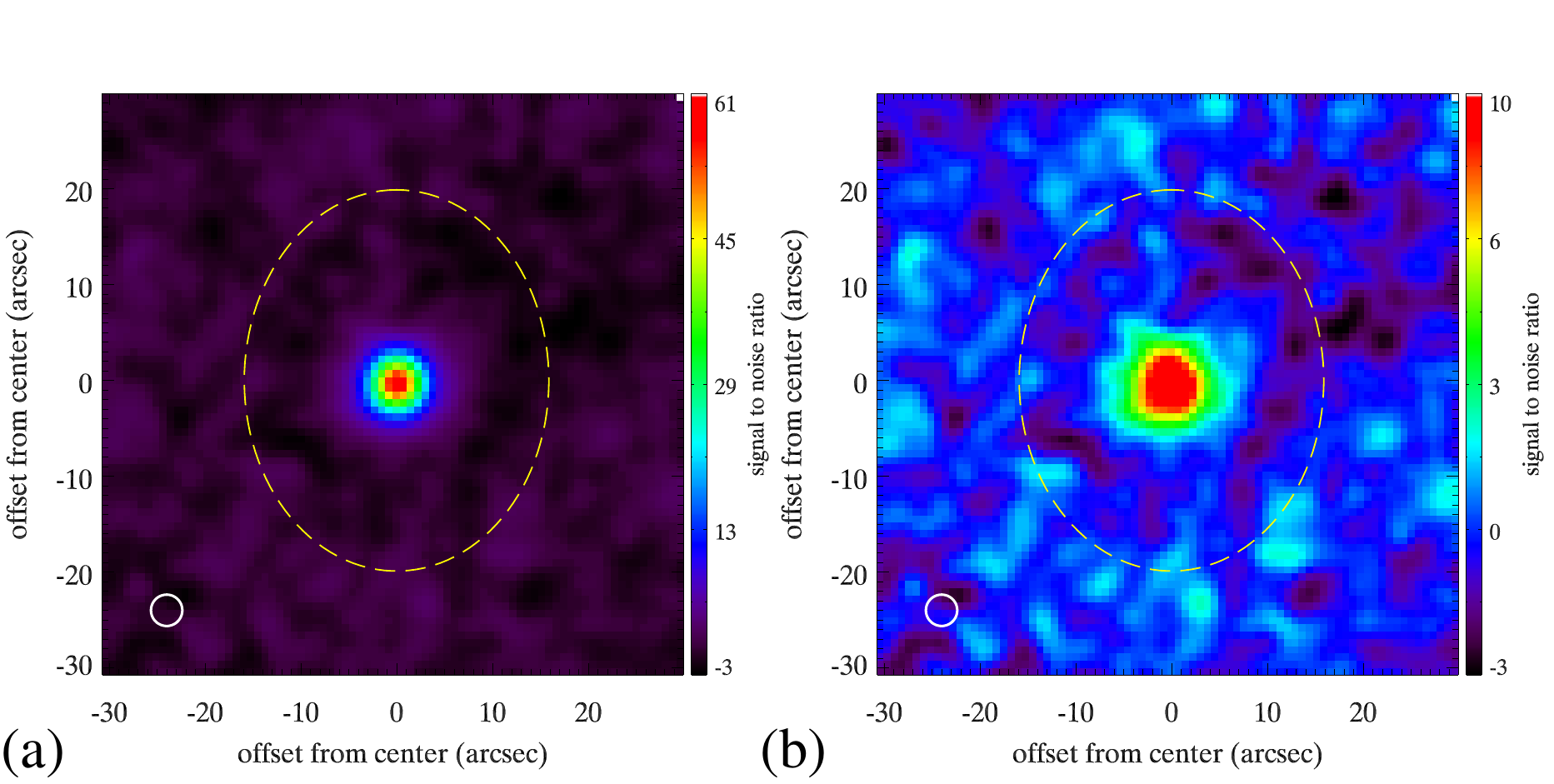}
  \caption{SOFIA 34.8 \um image of the $\epsilon$ Eri system. Both
images were smoothed by a Gaussian kernel of 1.5 pixels. The color
scale is shown in units of signal-to-noise ratio with 1 $\sigma$ of
0.6 mJy/arcsec$^2$. The white circle in the left corner in both panels
shows the beam size of the F348 filter, while the dashed yellow
ellipse marks the inclined 64-au cold Kuiper-belt analog.  The left
panel (a) shows the final coadded mosaics before stellar subtraction
while the right panel (b) shows after.}
\end{figure*}

Compared to the profile of the stellar calibrators (i.e., the hybrid
PSF), the \epseri profile is slightly extended in the range of
4\arcsec--10\arcsec\ from the star\footnote{Note that this range is
where there is a discrepancy between the observed and theoretical
PSFs. Since our hybrid PSF used observed PSFs within 10$''$, the
extension in the $\epsilon$ Eri data is real, not subject to PSF
uncertainty. Although we discounted the Ceres data from PSF comparison
because of possible non-sidereal tracking errors creating a larger PSF
than the stellar PSF, we do note that the \epseri profile from
3\arcsec--5\arcsec\ is still significantly larger than even the Ceres
profile.}. The slight extension is mostly evident at radii of
3\arcsec--5\arcsec\ (10--16 au at the distance of $\epsilon$ Eri), and
is independent of the registration methods. For simplicity, we adopt
the centroiding coadded image for further analysis. The final coadded
image of the $\epsilon$ Eri system is shown in Figure
\ref{epsEri_forcast}. The stellar photosphere of $\epsilon$ Eri is
estimated to be 0.81 Jy at 34.8 $\mu$m (Section 4.1). The total flux
within 10\arcsec\ is 1.30 $\pm$ 0.09 Jy, suggesting that we detect an
inner ($<$30 au) excess that is slightly more extended than the PSF.

\section{Analysis}

To quantify the amount and structure of the excess emission, the
stellar contribution needs to be subtracted from the image. To aid in
characterizing the inner 25 au of the debris structure in the
$\epsilon$ Eri system, we also include a re-analysis of the IRS
spectrum and MIPS 24 \um image of the system (previously published in
\citet{backman09}). In Section 4.1, we derive the photospheric flux at
both bands. We then assess the excess emission by performing PSF
subtraction at 35 \um in Section 4.2 and at 24 \um in Section 4.3. The
re-analysis of the IRS spectrum is presented in Section 4.4 where we
demonstrate the mid-infrared excess is consistent with dust emission
with a temperature of 150$\pm$20 K.

\subsection{Photospheric Fluxes}

We estimated the photospheric output of $\epsilon$ Eri as
follows. Since the star is only slightly cooler than the Sun, we used
the carefully determined SED of the Sun as a starting point
\citep{rieke08}. We took the effective temperature of the Sun to be
5780 K and adjusted the overall shape of the assumed SED of $\epsilon$
Eri by the ratio of blackbodies, one at the temperature of the Sun and
the other at a temperature assigned for $\epsilon$ Eri. We left the
latter temperature as a free parameter and varied it to minimize
$\chi^2$ as determined relative to photometry of the star at $H$,
$K_S$, {\it Spitzer }IRAC1, IRAC3, IRAC4, {\it WISE }W3, and
W4\footnote{Because of its known infrared excess, we could not use
direct measurements of $\epsilon$ Eri for the W3 and W4 photometry,
but instead we based the photospheric values on the color differences
with $K_S$ for the solar clones in \citet{gray06} and all K1 -- K3
dwarfs listed by \citet{gray03,gray06}, basing this calculation only
on the stars so described with accurate 2MASS photometry. By using
color differences in identical bands, we were able to circumvent
systematic errors associated with photometric bandpass corrections
(since the spectra of both types of star over this entire wavelength
range are to first order Rayleigh-Jeans).}. We omitted the IRAC2 band
because it contains the CO fundamental absorption, which we expect to
be deeper in $\epsilon$ Eri than in the Sun. We found a sharp minimum
in $\chi^{\mathrm{2}}$ at a temperature of 5127 K, which is in
satisfactory agreement with the nominal temperature for a K2V star
(the type of $\epsilon$ Eri, \citealt{difolco04}) of 5090K. The
photospheric flux densities indicated by this procedure are 0.81 Jy at
34.8 $\mu$m (FORCAST F348 band), and 1.74 Jy at 23.68 $\mu$m (MIPS 24
$\mu$m band).

\subsection{Excess Emission at 35 \um}

\begin{figure}
  \figurenum{4}
  \label{psfsub_hybrid} 
  \plotone{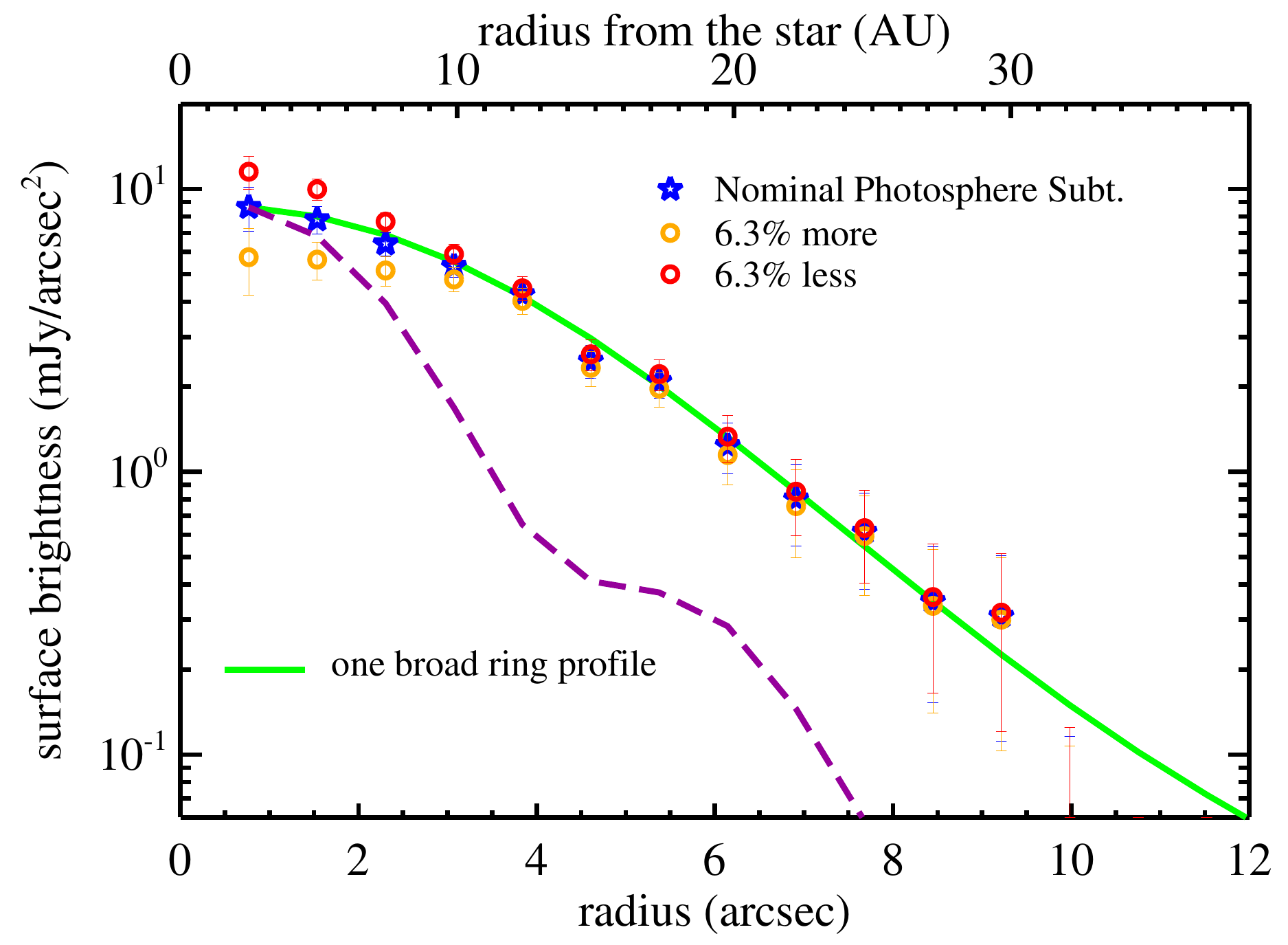}
  \caption{Excess-only profiles of the $\epsilon$ Eri system at 34.8
\um after photospheric subtraction (nominal and $\pm$6.3\% the
photospheric values. The PSF profile (normalized to the peak) is shown
as the dashed line. The excess emission is resolved at 34.8 \mm, which
could arise from (1) a central broad ring with a width of $\sim$18 au,
shown as the green line, or (2) one point source plus a ring peaked
at 10 au (not shown, see details in Section 4.2). }
\end{figure}

To characterize the excess emission at 34.8 $\mu$m near the star, PSF
subtraction is necessary. We scaled the hybrid PSF to match the
photosphere of $\epsilon$ Eri by normalizing its total flux within an
aperture of 12\arcsec\ to be 0.81 Jy without sky annulus (the sky is
zero in the hybrid PSF). To account for the absolute flux calibration
uncertainty, which includes (1) the photospheric prediction, 2\%, and
(2) the FORCAST flux calibration 6\%\footnote{The flux calibration
errors are given in the data headers and are a product of the {\it
SOFIA} Data Cycle System pipeline, which provides the calibration.},
the PSF subtraction was also performed after scaling the hybrid PSF
within $\pm$6.3\% of the nominal photospheric value, allowing us to
set the lower and upper boundaries of the uncertainty in the
PSF-subtracted image. The nominal photospheric subtracted image is
shown in Figure \ref{epsEri_forcast}b, and the excess-only radial
profiles are shown in Figure \ref{psfsub_hybrid}. The resultant peak
flux in the excess-only profiles varies by 30--40\%, depending
sensitively on the exact scales of the PSF fluxes. However, these
differences decrease significantly for the region outside the FWHM of
the beam (i.e., outside a radius of 2\arcsec). We also evaluated the
impact from the variation of the PSF FWHM (8\%, see Section 2.2) in
the photospheric subtraction. Using a narrower PSF, the resultant disk
flux near the core is expected to be lower while the flux outside the
core region would be slightly higher. Using a broader PSF, the
resultant disk profile should have an opposite effect (i.e., higher
flux near the core and slightly lower flux outside the core). We
tested the changes by artificially broadening and sharpening the scaled
PSF by 8\%, and found that the resultant disk profiles are still within
the uncertainty boundary set by the absolute flux
calibration. Therefore, the extension (compared to the PSF profile)
beyond 2\arcsec is robust, not subject to the uncertainties in
absolute flux calibration nor to the PSF subtraction.

The excess emission is resolved at 34.8 $\mu$m by $\gtrsim$ 2 beam
widths (i.e., the emission region is extended beyond 10 au). The
excess flux at 34.8 $\mu$m within 10\arcsec\ is 0.49$\pm$0.09 Jy, 60\%
of the stellar photospheric output. The observed profile is consistent
with (1) a broad Gaussian structure peaked at the star, with a width
of 18 au (green line in Figure \ref{psfsub_hybrid}), or (2) an
unresolved source at the center plus a Gaussian-profile ring peaked at
10 au, with a width of 10 au. In summary, the {\it SOFIA }data confirm
the excess emission near the star within 20 au, but cannot
differentiate whether the emission region is one broad ring or
composed of two separate structures (e.g., an unresolved source plus a
ring).

\subsection{Excess Emission at 24 \um}

We searched the {\it Spitzer } archive and found unpublished MIPS 24
$\mu$m data (AOR 8969984), which account for an additional 50\% of
integration depth in addition to the published data (AOR 4888832)
presented in \citet{backman09}. We used the MIPS instrument team
in-house pipeline \citep{gordon05,engelbracht07} to reprocess these
data to correct for instrument artifacts, and combined all data into a
final mosaic. The combined data appear to be point-like, but with a
FWHM of 5$\farcs$71$\times$5$\farcs$63, slightly extended compared
with a typical point source (5$\farcs$50$\times$5$\farcs$42). We
subtracted the photospheric contribution by scaling a blue calibration
PSF to the estimated photospheric value (1.74 Jy). The photospheric
subtracted image has a FWHM of 7$\farcs$17$\times$6$\farcs$89, much
broader than that of a point source.

\begin{figure}[t!] 
  \figurenum{5}
  \label{obs_radf_mips24} 
  \plotone{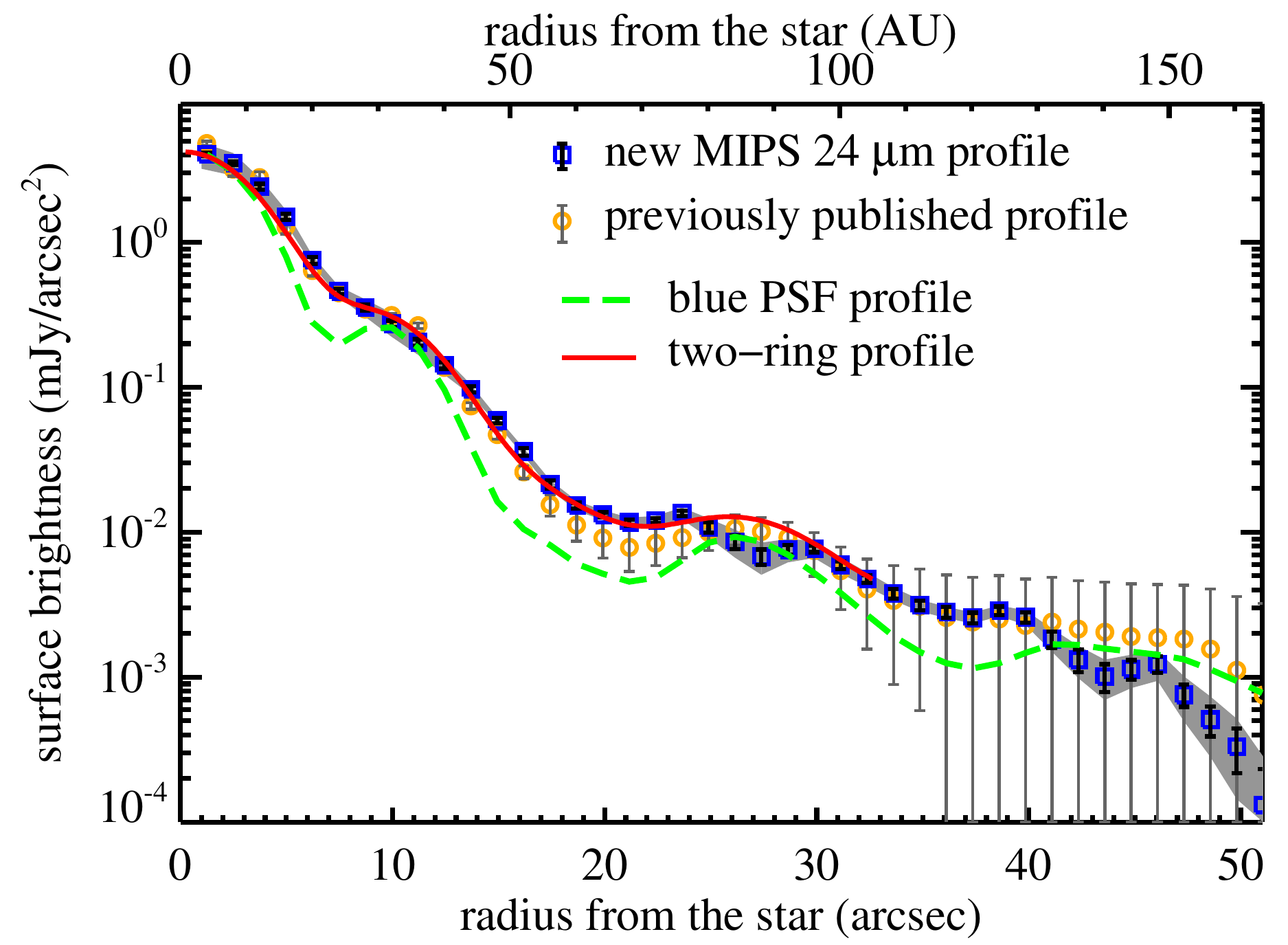}
  \caption{Excess-only profiles of the $\epsilon$ Eri system at 24 \um
after photospheric subtraction. The profile using the nominal
photospheric value is shown as color symbols with circles for the new
one produced in this work in comparison with the published profile
(squares). The gray area around the new profile marks the uncertainty
boundary due to photospheric subtraction and absolution flux
calibration ($\pm$2.8\% the photospheric values). The PSF profile
(normalized to the peak) is shown as the solid green line. The first
and second dark Airy rings are partially filled, suggesting the excess
emission is slightly resolved at 24 \mm. The red line shows the
resultant radial profile for a two-ring model (for details see Section
4.3).}
\end{figure}

To characterize the excess emission at 24 $\mu$m further, we computed
the azimuthally averaged radial profiles of the excess emission, shown
in Figure \ref{obs_radf_mips24}. As at 35 $\mu$m, the uncertainty in
the photospheric profiles was estimated by repeating the analysis with
adjustments of $\pm$2.8\% in the nominal photospheric value (2\% from
the absolute flux calibration and 2\% from the photospheric
extrapolation). The new 24 $\mu$m surface brightness profile is very
similar to the one published by \citet{backman09} (Figure
\ref{obs_radf_mips24}), but the improved reduction substantially
reduces the errors at larger radii (enhanced by 1/$\sqrt{N}$ where N
is the number of pixels in each of the annuli). Compared to the
profile of a point source, the photospheric subtracted image has the
first and second dark Airy rings (corresponding to 20 au and 65 au in
$\epsilon$ Eri) partially filled, supporting the evidence that the
excess emission is slightly resolved in the MIPS 24 $\mu$m band.

To gain insights into the spatial distribution of the excess emission,
we constructed a simple geometric two-ring model. The first ring is
fixed at the star position with a specified width and represents the
emission inside 20 au, and the second ring represents the emission
from the cold Kuiper-belt-like ring, with a specified peak position and
width. A final, high-resolution synthesized image is the sum of these
two rings with a given relative flux ratio and a fixed total flux;
i.e., there are four free parameters in this two-ring model: the width
of the central ring, the peak position of the outer ring, the width of
the outer ring, and the relative flux ratio. These high-resolution
model images were then inclined to view at 30\arcdeg\ from face-on,
and convolved with the 24 $\mu$m PSF to simulate the observations.

We varied the four model parameters to minimize $\chi^2$ relative to
the observed radial profile within 25\arcsec. We found that a central
ring with a radius of $\sim$13 au and an outer ring peaked at 64 au
with a width of $\sim$24 au can fit the observed radial profile
relatively well (red line in Figure \ref{obs_radf_mips24}). It is
reassuring that the peak position of the outer ring is found to be
similar to the location found in other studies ($\sim$64 au,
\citet{backman09,macgregor15,chavez16} ). However, as in the 35 $\mu$m
profile analysis, this does not indicate there is only one inner
excess region, but only that the inner excess emission region is
extended at least to $\sim$13 au.

\begin{figure}[thb] 
  \figurenum{6}
  \label{obs_irs} 
  \plotone{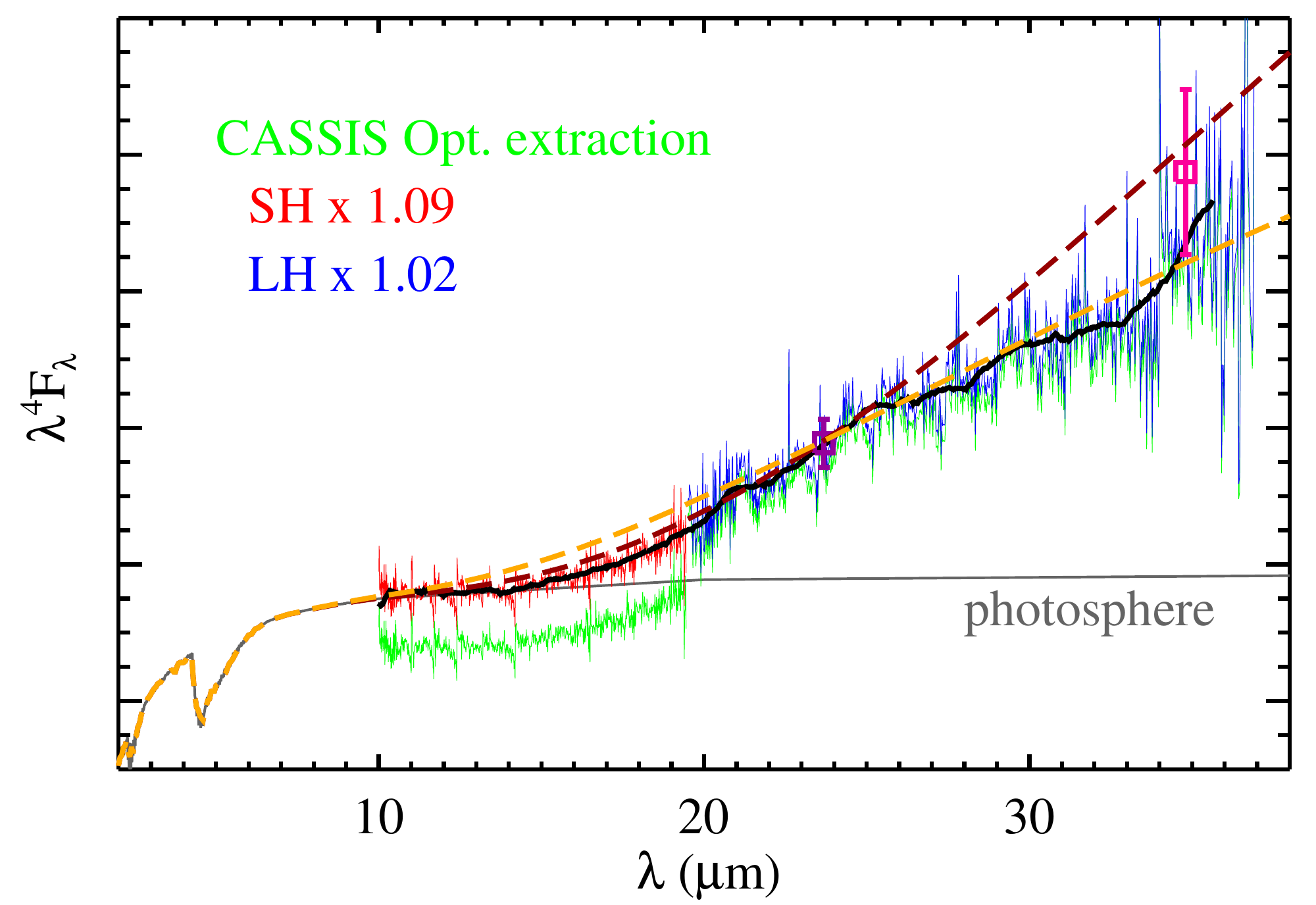}
  \caption{{\it Spitzer} IRS high-resolution spectrum of $\epsilon$
Eri in $\lambda^4F_{\lambda}$ vs.\ $\lambda$ format so a
Rayleigh-Jeans spectrum is flat. The MIPS 24 and FORCAST 35 \um
photometry is shown as squares for reference. The green color shows
the CASSIS optimal extraction spectrum. The red and blue color lines
are the scaled SH and LH modules, respectively (see Section 4.4 for
details). The final combined (joined and smoothed) spectrum is shown
as the thick black line. The two dashed lines are the sum of the
photosphere and a blackbody emission of 130 K (orange color) and 170 K
(brown color). The excess emission is consistent with a blackbody
emission of $\sim$150$\pm$20 K.}
\end{figure}

\subsection{{\it Spitzer} IRS High-Resolution Spectrum}

Details about the {\it Spitzer} IRS observations were given in
\citet{backman09}. As stated in that paper, linearity and saturation
are an issue in the IRS low-resolution data. Therefore, we only
discuss the high-resolution data (SH: 9.9--19.5 \um with a slit size
of 4\farcs7$\times$11\farcs3 and LH: 18.7--37.2 \um with a slit of
11\farcs1$\times$22\farcs3). We retrieved the extracted high
resolution spectrum from the CASSIS website\footnote{The Combined
Atlas of Sources with Spitzer IRS Spectra (CASSIS) is a product of the
IRS instrument team, supported by NASA and
JPL. http://cassis.sirtf.com/atlas/cgi/browse-hires.py} and used the
``optimal'' product which simultaneously determines the source
position and extraction in the two different nod observations.  As
described in \citet{lebouteiller15}, this mode of extraction produces
the best results when the source is resolved but only marginally
extended. The CASSIS spectrum is shown in Figure \ref{obs_irs} in the
$\lambda^4F_{\lambda}$ vs.\ $\lambda$ format so a Rayleigh-Jeans
spectrum is flat. There is a flux jump between the two SH and LH
modules: the SH part of the spectrum is lower than the expected
photosphere determined in Section 4.1, and the LH part of the spectrum
is slightly lower than the MIPS 24 \um photometry. We joined the two
modules by (1) scaling the SH module by 1.09 so that the 10--12 \um
region matches the expected photosphere level, and (2) scaling the LH
module by 1.02 so it matches the MIPS 24 \um photometry. The scalings
are within the uncertainty in the absolute flux calibration between
the MIPS and IRS instruments.

We used the final combined and smoothed (to R = $\lambda/\Delta
\lambda \sim$30) spectrum (the black line in Figure \ref{obs_irs}) to
estimate the dust temperature of the excess emission. Although the excess
is not exactly blackbody-like, the excess emission can be described by 
a blackbody emission  with temperatures between 170 K and 130 K. In Figure
\ref{obs_irs}, we overplotted two (dashed) curves to represent
the sum of the photosphere and a blackbody emission of 170 K and 130 K
(both normalized to the MIPS 24 \um photometry point). This comparison
suggests that the excess emission is consistent with a dust
temperature of $\sim$150$\pm$20 K.

\begin{figure}[hbt] 
  \figurenum{7}
  \epsscale{1.2}
  \label{dust_td} 
  \plotone{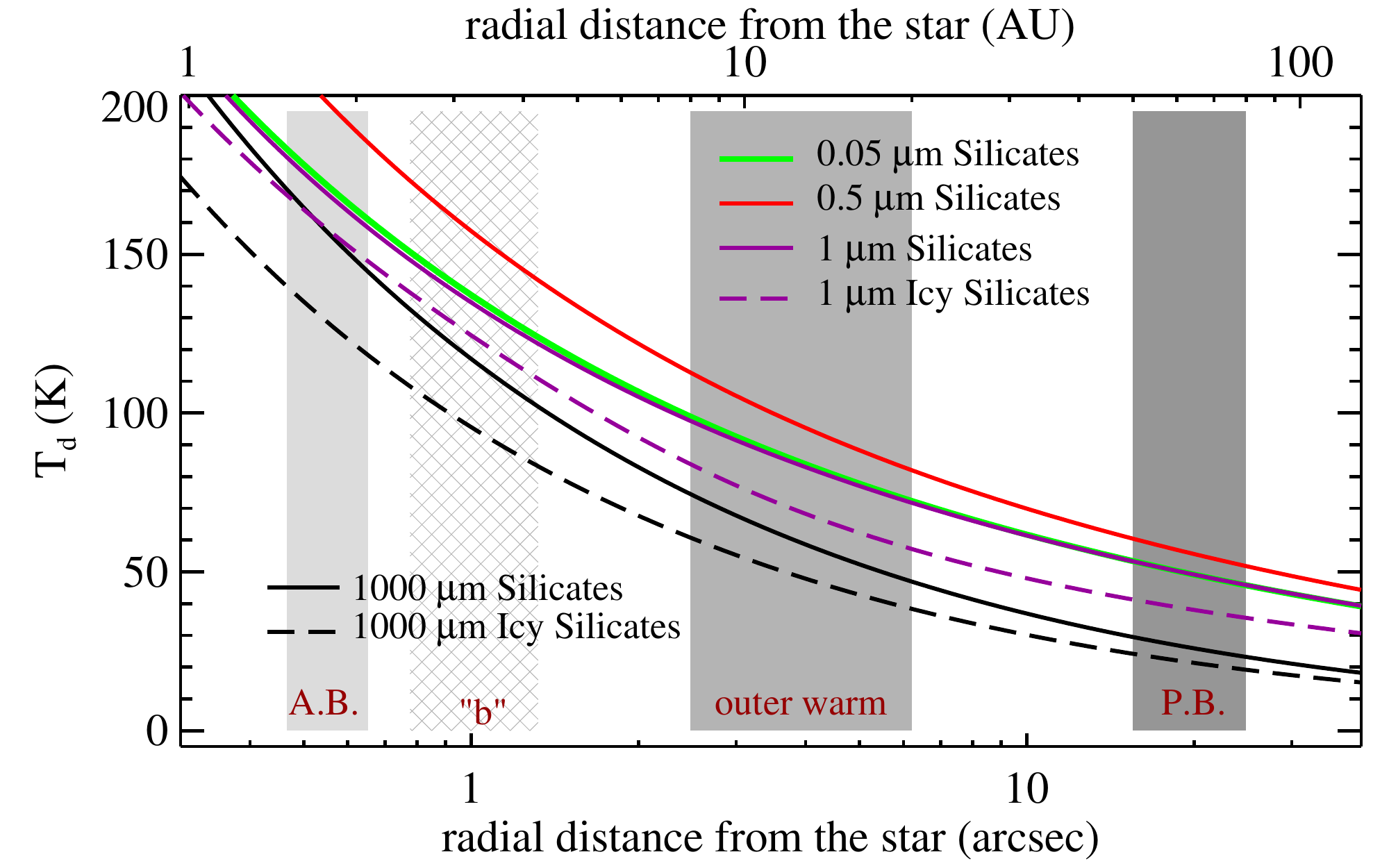}
  \caption{Dust temperature distribution for both astronomical
silicates and icy silicates for selected grain sizes around $\epsilon$
Eri. The chaotic zone of the putative \epseri b is marked as the hashed
area. The two light gray areas represent the two distinct dust belts
(asteroid belt (A.B.) and outer warm belt) that are consistent with
the presence of \epseri b (see Section 5.3 for details). The dark gray
area marks the Kuiper-belt-like planetesimal belt (P.B.).}
  \epsscale{1.0}
\end{figure}

For blackbody-like emitters (1 mm astronomical silicates), these
temperatures correspond a stellocentric distance of 1.5--2.5 au;
however, for 1 \um silicate-like grains the corresponding distance is
larger ($\sim$2--3.5 au) (see Figure \ref{dust_td}). This
temperature-radius relation is also composition dependent. In Figure
\ref{dust_td} we show the temperature distributions using icy
silicates (90\% of ice by volume). Icy grains are generally poor
absorbers; therefore, at the same temperature they are located at
smaller stellocentric distances compared to the same size, bare
silicates.

\section{Model Comparison}

Based on the analysis presented in Section 4, it is evident that there
is a substantial amount of excess emission in the inner 25 au of the
$\epsilon$ Eri system, which is resolved by FORCAST at 35 $\mu$m and
MIPS at 24 \um (at a linear resolution of $\sim$ 11 au). We refer to
this excess emission as the ``warm'' excess to differentiate it from
the cold excess emission from the 64 au Kuiper-belt-like ring. Two
different scenarios have been suggested for the origin of the warm
excess around $\epsilon$ Eri: (1) in-situ planetesimal belt(s) and (2)
grains dragged in from the cold Kuiper-belt-like belt. Using the newly
obtained mid-infrared disk radial profiles, we test these proposed
models in the following subsections to probe the nature of the warm
excess around this star.

We test the models proposed by \citet{backman09},
\citet{reidemeister11} and \citet{greaves14} for the inner 25 au
region. High-resolution face-on model images at 23.68 $\mu$m and 34.8
$\mu$m are constructed based on the parameters given in those
papers. Each of the models also reproduces the system's SED globally
when additional components are added. We explore two different grain
properties: astronomical silicates \citep{laor93} and a mixture of
silicates and organics \citep{ballering16} to fit the mid-infrared
disk profiles. We find that the choice of the grain types and
properties has only a small impact on the resultant model images at
these wavelengths. Given the degeneracy, we only show the model
results using astronomical silicates. Furthermore, icy silicates are
used when computing the SEDs for the components beyond the ice line
(radial location $>$ 4 AU, e.g., the cold planetesimal belt and outer
warm belt). The high resolution, face-on model images are then
inclined to view at 30\arcdeg\ from face-on and convolved with
instrumental PSFs to simulate the observations. We compare the model
radial profiles with the observed ones for each case. Since we only
focus on the nature of the warm excess, the comparison was only done
for the disk surface brightness profile within $\sim$10\arcsec\ (30
au).

\subsection{Dragged-in Grains as Proposed by \citet{reidemeister11}}

The stellar wind drag for $\epsilon$ Eri is found to be 28 times
stronger than the P-R drag, based on the measured mass-loss rate 30
times higher than that of the Sun \citep{wood02} and the average solar
wind velocity. Therefore, a significant inward flow of dust grains
from the cold Kuiper-belt-like region is expected, unless there is a
massive, shepherding planet interior of the cold belt. Assuming no
such planet, \citet{reidemeister11} modeled the marginally resolved
{\it Spitzer }24, 70 and 160 $\mu$m images and found that the {\it
Spitzer} data are consistent with this possibility.

\begin{figure}[thb] 
  \figurenum{8}
  \label{model_drag-in} 
  \plotone{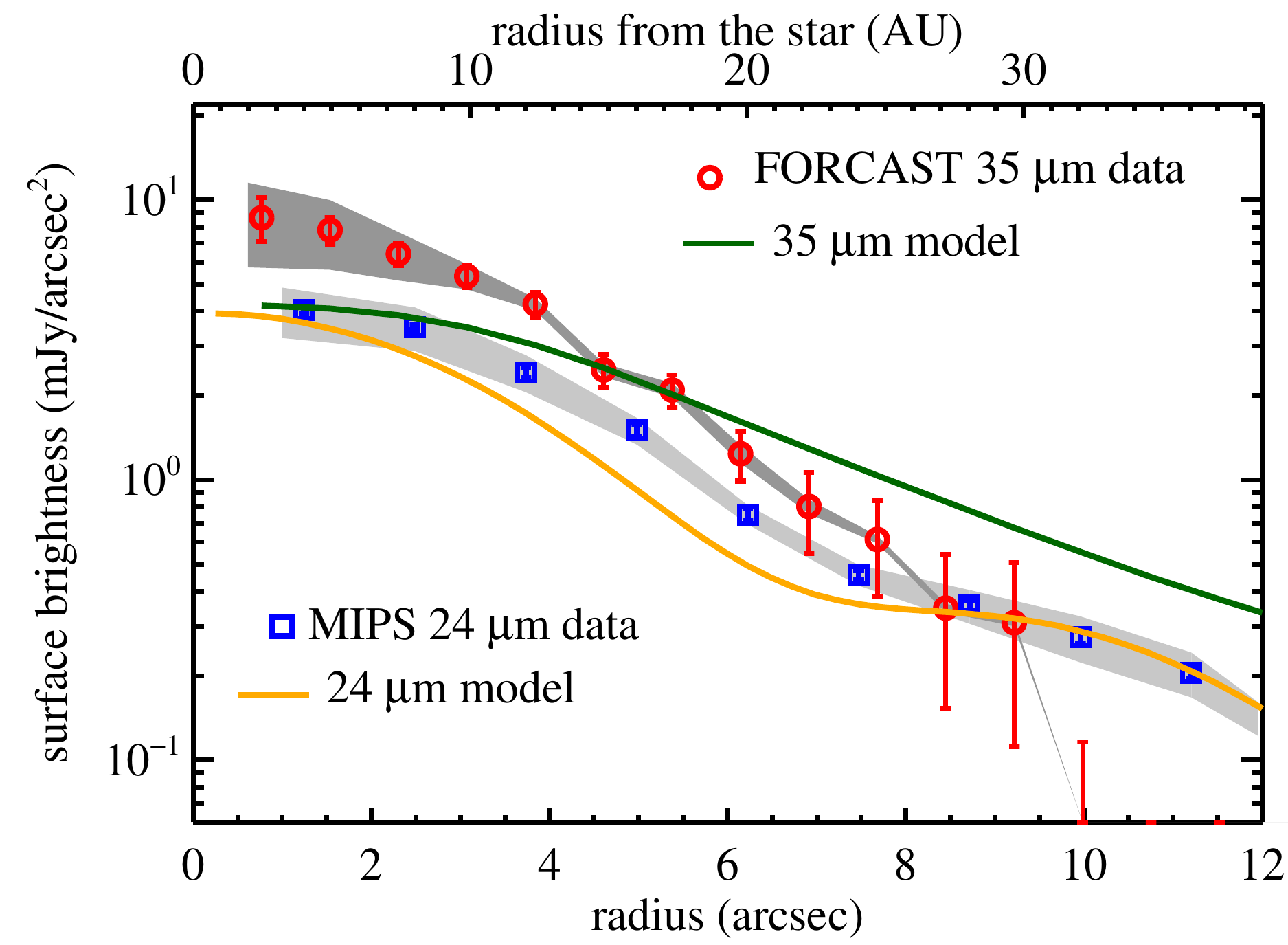}
  \plotone{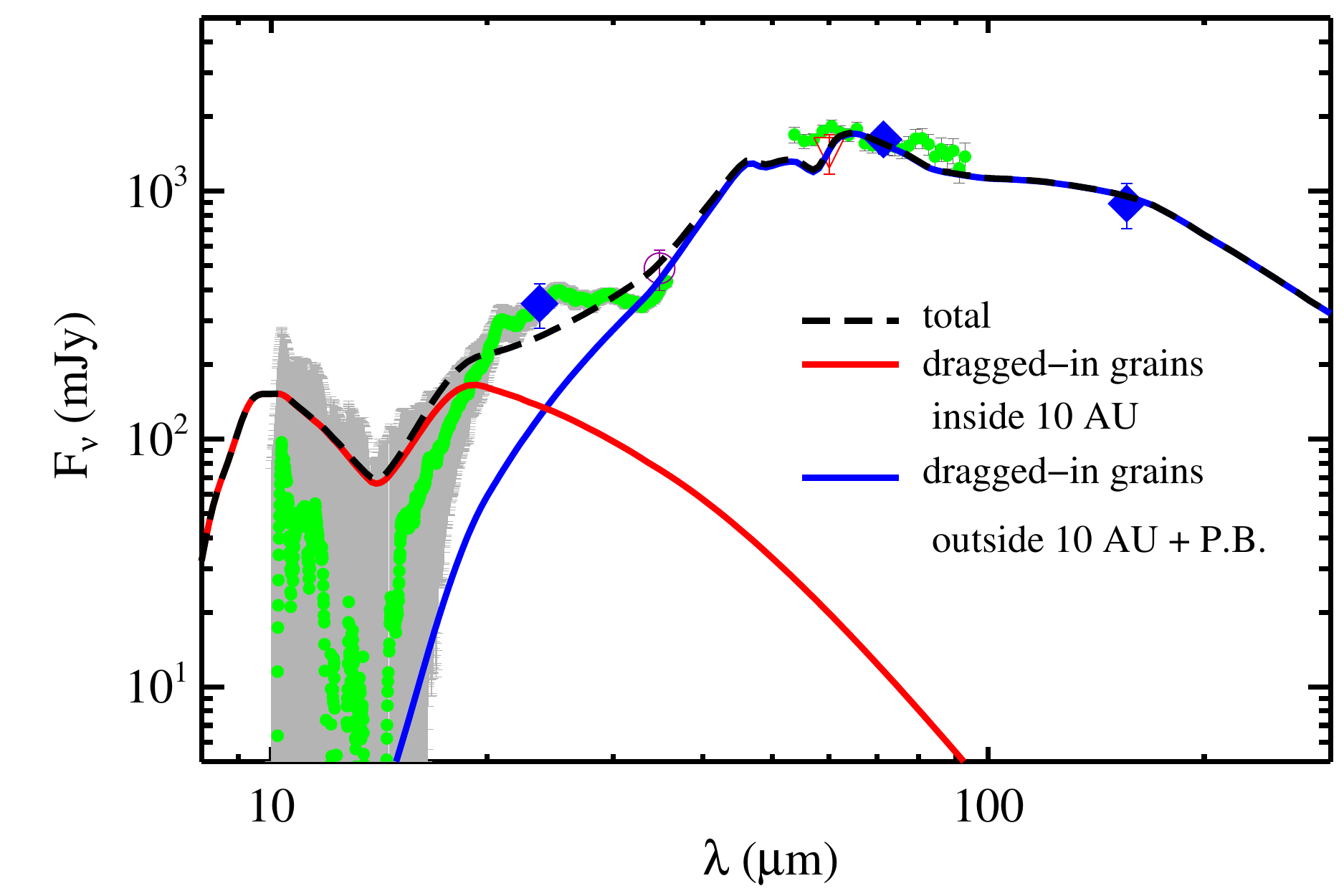}
  \caption{Model results using the parameters from
\citet{reidemeister11} in comparison with the observed disk profiles
(upper panel) and the SED (bottom panel).  In the disk profile plot,
the observed profiles are symbols with gray areas marking the upper
and lower boundaries of uncertainty due to flux calibration and PSF
subtraction. Red circles are the FORCAST 35 \um data, and blue squares
are the MIPS 24 \um data. Model profiles are shown in solid lines with
green color for 35 \um and orange for 24 \mm. In the bottom SED plot,
the various symbols are the excess photometry measurements for the whole
system, and the green dots with error bars are the excess from the IRS
and MIPS-SED spectra. The gray area around the IRS excess spectrum
represents the uncertainty associated with the stellar photospheric
extrapolation and subtraction.  The model SED is composed of two
parts: the red (the dragged-in small grains inside 10 au) and blue
solid lines (the rest of the dragged-in grains and the dust in the
cold planetesimal belt (P.B.)) with the sum shown as the black dash line
(replicated from the bottom panel of Figure 8 in
\citealt{reidemeister11}).  }
\end{figure}

Using the no-planet configuration (i.e., no dynamical perturbation in
the dragged-in dust flow), we computed model images with parameters
derived from \citet{reidemeister11} to compare with our new data. The
derived radial disk profiles are shown in the upper panel of Figure
\ref{model_drag-in} with the SED shown in the bottom panel. The model
SED is a replicate of the bottom panel of Figure 8 in
\citet{reidemeister11} using the same grain parameters and
composition. The model disk surface brightness profile at 24 $\mu$m
is consistent with the one published in \citet{reidemeister11} (i.e.,
the fit is good for the old, large error-bar profile). However, the 35
$\mu$m model disk profile under-predicts the disk flux within
4\arcsec\ (12 au), and over-predicts it outside 6\arcsec\ (20 au). The
proposal by \citet{reidemeister11} that the entire inner warm disk
could result from dragged-in grains is not consistent with the {\it
SOFIA } measurements.

\begin{figure}[thb] 
  \figurenum{9}
  \label{model_greaves} 
  \plotone{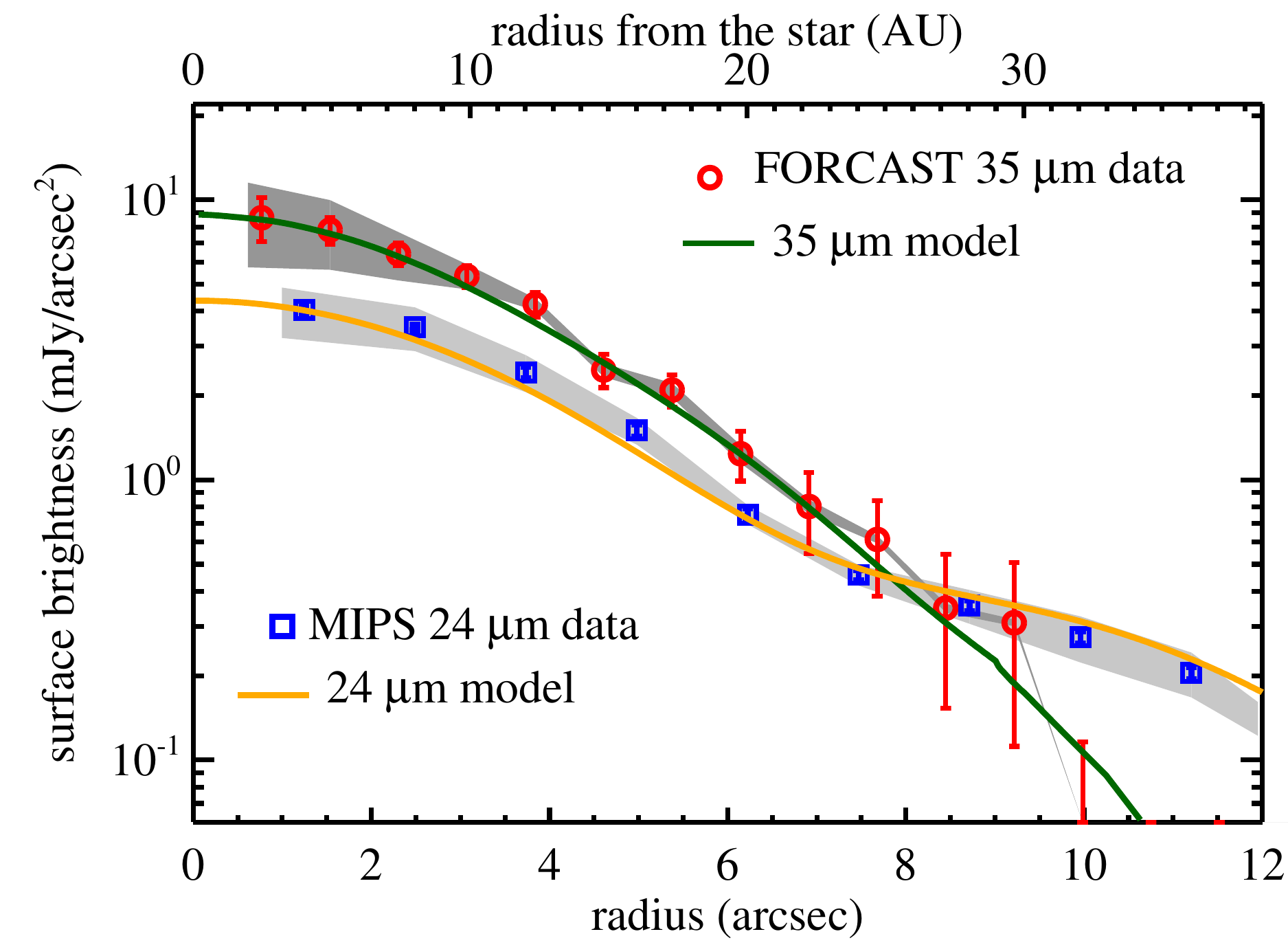}
  \plotone{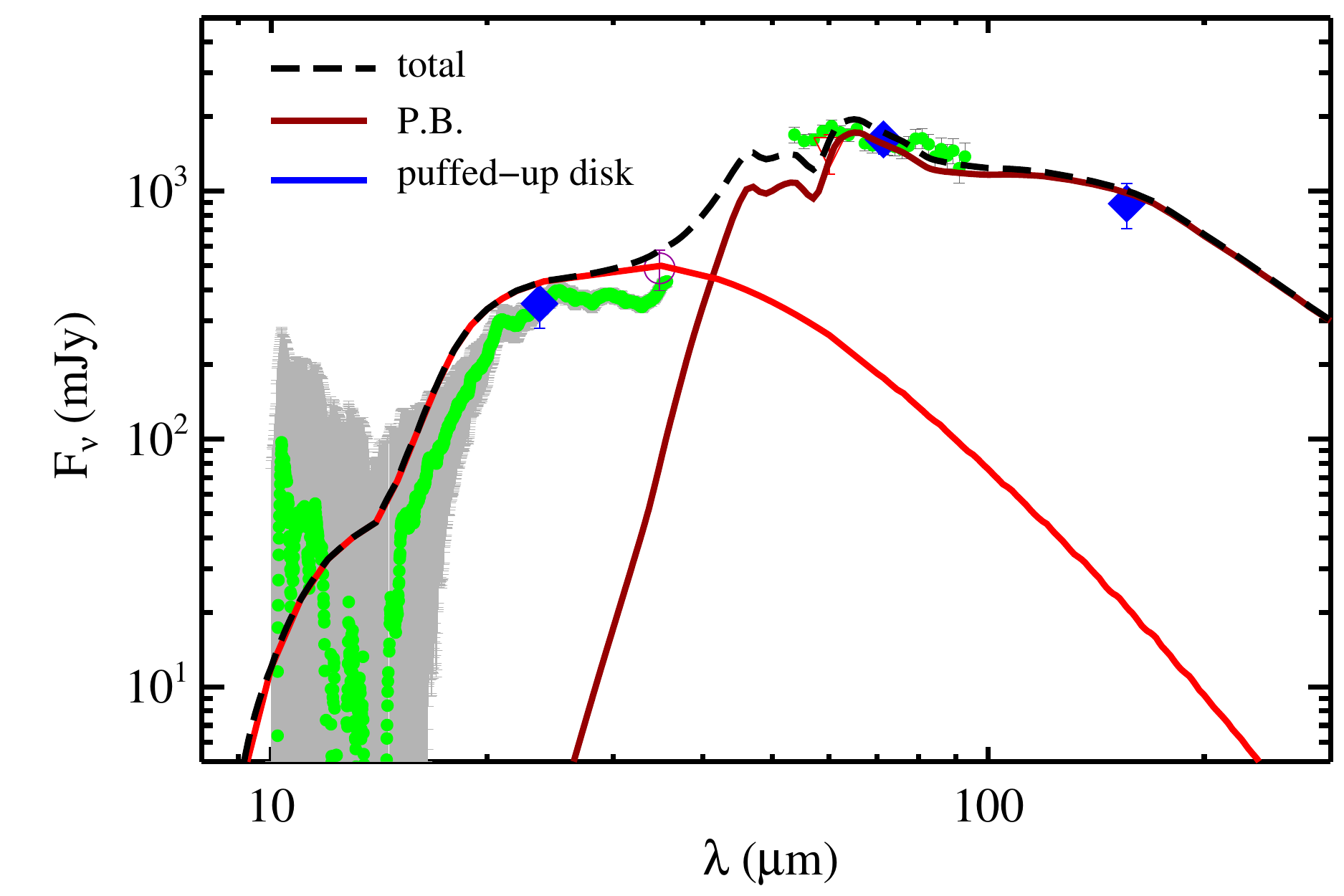}
  \caption{Model results using the geometric parameters of
\citet{greaves14} in comparison with the observed disk profile (upper
panel) and SED (lower panel). Symbols and lines are the same as in
Figure \ref{model_drag-in}. In the bottom SED plot, the model SED for
the puffed-up dust disk is shown as the red solid line and the broad
cold disk is shown as the brown solid line with the sum shown as 
the black dashed line. }
\end{figure}

\subsection{One Broad and Puffed-up In-situ Dust Belt as Proposed by \citet{greaves14}}

Resolved images of $\epsilon$ Eri obtained by {\it Herschel }suggest
inner excess emission at 70 and 160 $\mu$m. \citet{greaves14} modeled
this inner excess as one single disk with simple geometric parameters:
a wedged disk with a radial span of 3--21 au in an $r^{-1}$
density distribution and an opening angle of $\pm$23\arcdeg. Note
that the opening angle is quite large, i.e., the disk is vertically
extended. Due to the significant scale height in this model, we
computed the model images with the code {\it dustmap } ver.\ 3.1.2
\citep{stark11}, which can take a 3-D structure as an input, with the
geometric parameters listed above and an inclination angle of
30\arcdeg.  We assumed grain parameters using compact astronomical
silicates with a minimum grain size ($a_{min}$) of 1 $\mu$m, a maximum
grain size ($a_{max}$) of 1000 $\mu$m, and a particle size power-law
index ($q$) of --3.5. Figure \ref{model_greaves} shows the
results. This model fits the disk profiles very well (the upper
panel of Figure \ref{model_greaves}), and reproduces the mid-infrared
SED reasonably well with the normalization set by fitting the disk
profiles.  For the sake of completeness, we also computed the cold
belt SED using the parameters given by \citet{greaves14} -- a broad,
geometrically thin disk from 36 to 72 au with a constant surface
density. For this cold component, we used icy silicates with $a_{min}$
of 1 \um and $a_{max}$ of 1000 \um in a $q=-3.5$ size distribution. We
did not include this component when computing the 24 and 35 \um disk
profiles since its contribution is insignificant at these
wavelengths. The combined SED fits the observed points
satisfactorily. However, we note that the inner radius of this cold
disk (36 au) is significantly smaller than the one inferred from the
mm observations ($\sim$53 au from \citet{macgregor15}, and $\sim$59 au
from \citet{chavez16}). This might suggest that the broad, cold disk
geometry is too simplistic, and a more complex distribution (e.g., a
narrow cold belt plus an outer warm belt (see Section 5.3) or a small
amount of dragged-in grain component) is needed.

The choice of $a_{max }$ has no impact on disk profiles nor the SED at
the model wavelengths; however, the value of $a_{min}$ does make a
noticeable difference at 24 $\mu$m and in the overall shape of the SED
shortward of $\sim$20 $\mu$m. In general, the $a_{min}$ is usually set
to the radiation blowout size ($a_{bl} = 1.14\ Q_{pr}\ L_{\ast}\
M_{\ast}^{-1}\ \rho_g^{-1}$ where $\rho_g$ is the grain density and
$Q_{pr}$ is the radiation pressure efficiency averaged over the
stellar spectrum, and is assumed to be 1 for grain sizes comparable or
larger than the wavelength where the stellar spectrum peaks
\citep{burns79}). However, as noted by \citet{reidemeister11} (Figure
2 in that paper), for the luminosity and mass of $\epsilon$ Eri, the
blowout size does not exist. In the \citet{reidemeister11} model,
$a_{min}$ was formally set to be 0.05 \mm, but much of the dust cross
section in their modeled size distribution comes from larger,
$\mu$m-sized grains (see their Figure 4). Setting $a_{min}$ to be 0.05
\um initially, we found that it is difficult to obtain simultaneous
good fits to the disk profiles at both wavelengths; i.e., a good fit
at 35 \um produces a too bright and broad MIPS 24 \um profile. Setting
$a_{min}$ to larger sizes significantly improves the fit at 24 \um as
shown in Figure \ref{model_greaves}. Setting different $a_{min}$ for
the cold broad disk has no noticeable difference in the resultant SED.
We will discuss the physical reason why a large $a_{min}$ is preferred
around $\epsilon$ Eri in Section 6.1.

\subsection{Two Narrow, In-situ Planetesimal Belts}

The third model we tested is similar to the two-belt model proposed by
\citet{backman09} based on the marginally resolved {\it Spitzer}
images. As noted by \citet{backman09}, the exact locations of the warm
belts ($\sim$3 au and $\sim$20 au) are not well constrained by either
the images or the SED. As suggested by the {\it Herschel }
measurements, the outer warm component is likely to be smaller than
was inferred from the {\it Spitzer } data. \citet{greaves14} suggest
the warm planetesimal belt is located at 14 au with a width of 4
au\footnote{although this is very different from the geometric model
given in the paper derived by fitting the PACS 70 and 160 $\mu$m
profiles.}. If the inner warm belt were a direct analog of our own
Asteroid belt (near the ice line), the expected location should be
1.5--2 au simply from scaling by stellar luminosity. In fact, such a
small size for the inner warm belt would be more consistent with the
presence of $\epsilon$ Eri b (see discussion in Section 6.2). 

In light of these results, we construct a revised two-belt model with
the following parameters. Both belts are assumed to be geometrically
thin (no scale height) and with a constant surface density, and to
contain grains from $a_{min}=$ 1 $\mu$m to $a_{max}=$ 1000 $\mu$m in a
power-law size distribution with $q$ set to --3.65 (e.g.,
\citealt{gaspar12}). The inner warm belt is assumed to range from
1.5--2 au, and the outer warm belt ranges from 8--20 au. Note that the
outer warm belt appears to be broad, but most of the emission at the
model wavelengths comes from the inner 8--12 au region due to radial
dependence of the dust temperatures since we used a constant surface
density. Furthermore, in order to produce overall good fit to the
system's SED, we have to use icy silicates for the outer warm belt to
suppress the total flux in the mid-infrared while providing enough
flux in the range of MIPS-SED spectrum (also see Figure 6 in
\citet{reidemeister11}). Similarly, we also include a SED fit for the
cold planetesimal disk ranging from 55 to 80 au (the radial span
derived by \citet{macgregor15}) using icy silicates with the same
grain size distribution as in the warm belts.

\begin{figure}[thb] 
  \figurenum{10}
  \label{model_twobelts} 
  \plotone{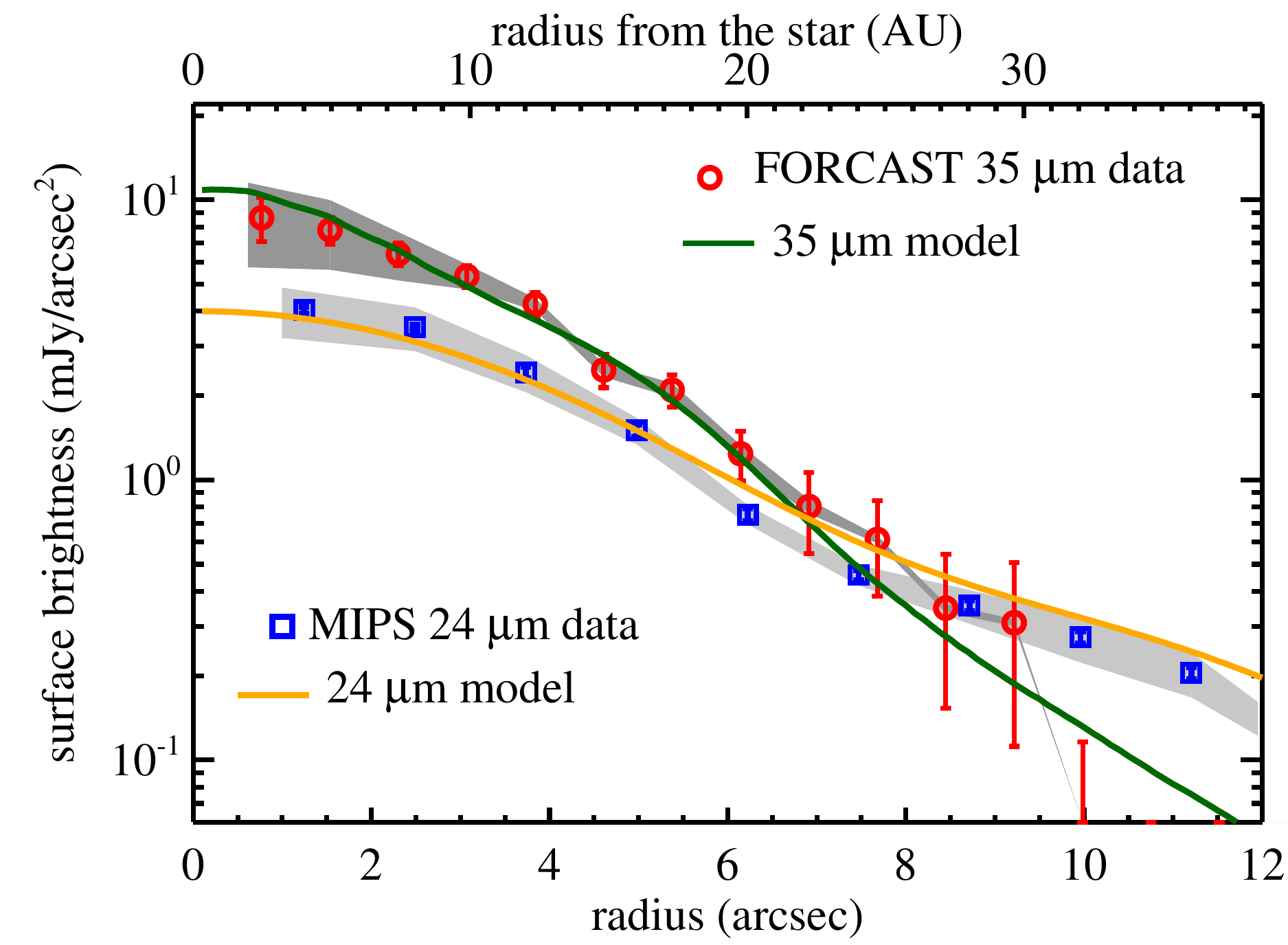}
  \plotone{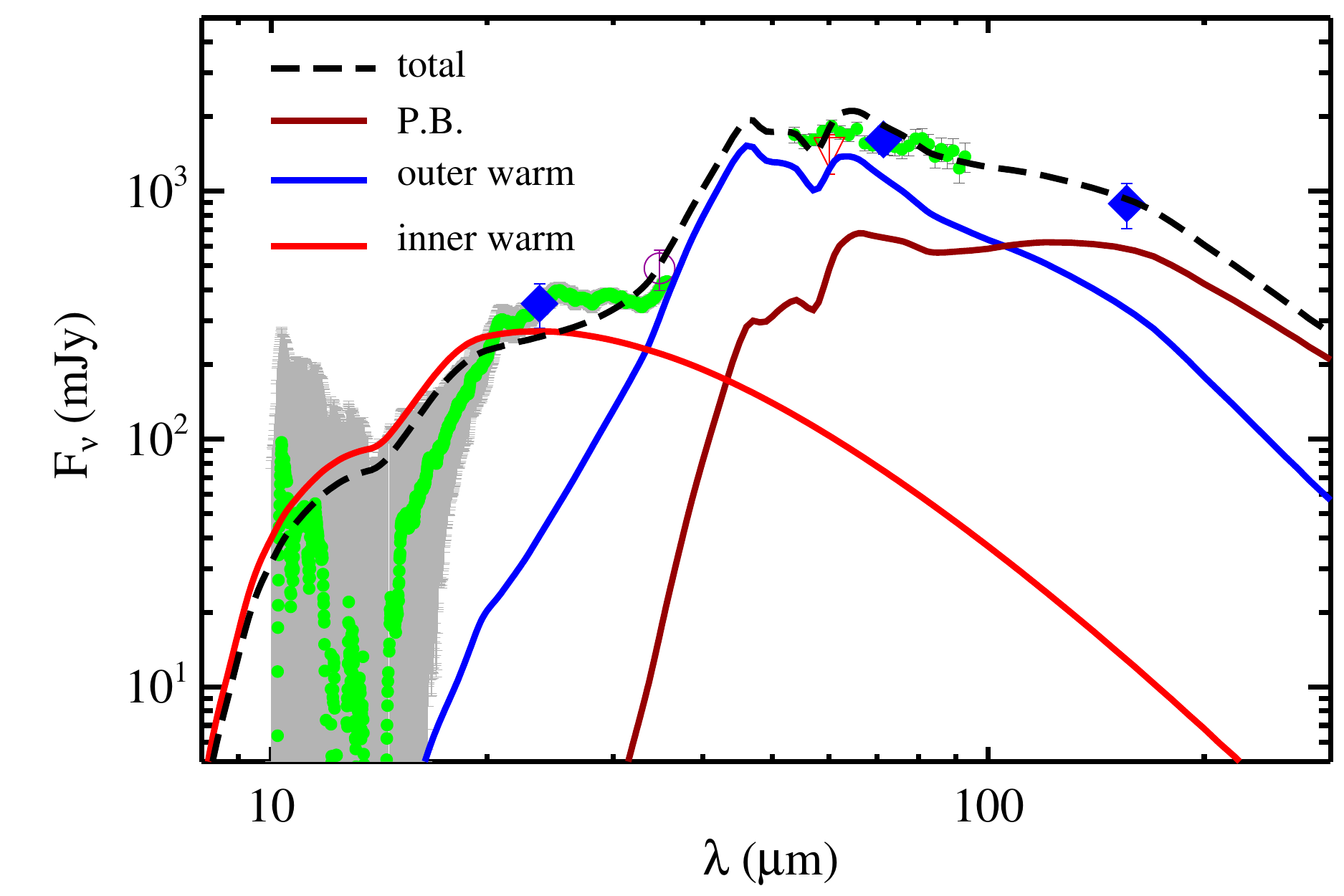}
  \caption{Model results using the revised two-belt model with the
locations slightly different from the ones of
\citet{backman09}. Symbols and lines are the same as in Figure
\ref{model_drag-in}. In the bottom SED plot, the two warm belts are
shown as the red and blue solid lines, and the contribution from the
dust in the 64 au belt is shown as the brown solid line. Similarly,
the sum of these three components is shown as the black dashed line. }
\end{figure}

The resultant model profiles are shown in the upper panel of Figure
\ref{model_twobelts} with the combined SED in the bottom panel. This
revised two-belt model fits the 35 $\mu$m profile reasonably well
(within 1$\sigma$), but is slightly too broad/bright outside 5\arcsec\
at 24 $\mu$m (but still within 3$\sigma$). The largest issue with
models of this class is that they fall slightly short of the measured
SED between 20 and 30 $\mu$m, which might indicate that the assumed
geometry is too simple, or might be revealing a problem for this type
of model. We also produced the model fits for a larger (3--4 au)
asteroid belt as originally suggested by \citet{backman09} with the
same outer warm belt (3--20 au) and SED parameters. The results are
very similar in the disk profiles with a slightly better SED fit in
the 20--30 \um region. Limited by the uncertainty in the exact shape
of the mid-infrared excess, both asteroid-belt models produce similar
results.

\subsection{Summary} 

We tested three proposed debris distributions in the inner 25 au of
the $\epsilon$ Eri system using the newly obtained mid-infrared disk
profiles. We found that the 24 and 35 $\mu$m emission is consistent
with the in-situ dust distribution produced either by one planetesimal
belt at 3--21 au (e.g., \citealt{greaves14}) or by two planetesimal
belts at 1.5--2 au (or 3--4 au) and 8--20 au (e.g., a slightly
modified form of the proposal in \citealt{backman09}). The observed
profiles are not consistent with the case dominated by dragged-in
grains (uninterrupted dust flow from the cold Kuiper-belt-analog
region) as proposed by \citet{reidemeister11}. This might suggest the
need of a planet interior to the 64-au cold belt to maintain the inner
dust-free zone, or a very dense cold belt where the intense collisions
destroy the dust grains before they have enough time to be dragged
in. In either case, some amount of dragged-in grains from the cold
belt can still contribute a fraction of the emission inside 25 au; the
exact amount remains to be determined by future high spatial
resolution data and improved collisional models for the cold belt. 

The model derived dust fractional luminosity ($f_d$) is
3$\times10^{-5}$ and 7$\times10^{-5}$ for the inner and outer warm
belts, respectively, in the revised two-belt model. For the broad disk
model, the dust fractional luminosity is 6$\times10^{-5}$. We adopt
the simple analytical model proposed by \citet{wyatt07} to test
whether the dust in the inner \epseri system is produced by transient
events. Assuming the typical parameters for disks around solar-like
stars and an age of 800 Myr, the maximum dust fractional luminosity
($f_{max}$) is 1$\times10^{-6}$ and 4$\times10^{-5}$ for a belt at 2
and 10 au. Therefore, the observed dust levels in these belts are
close to the expected maximum value for dust being generated through
collisional griding, and do not require to invoking transient
events\footnote{In the \citet{wyatt07} model, only the systems with
$f_d >> 1000 f_{max}$ are considered to be undergoing transient
events.}.

We note that the model parameters (especially the grain parameters)
are not unique in our test cases due to degeneracy between grain
properties and dust location. This is particularly true for the cold
belt component since we do not include the image fits to the
far-infrared and mm wavelength data (beyond the scope of the
paper). For simplicity, we adopted the astronomical silicates as
the grain composition in modeling the component inside 10 au. The
mismatch in the 20--30 \um SED region might partially be due to this
choice, in addition to the simple geometry assumed for the inner component. 
Furthermore, a larger (3--4 au) inner warm belt in the two-belt model
produces similar results. Nevertheless, the existence of an inner
(within 25 au) separate dust source (i.e., planetesimal belt(s)
different from the outer cold belt) is a robust conclusion inferred
from the newly obtained {\it SOFIA }data.

\section{Discussion} 

\subsection{Minimum Grain Sizes in the Inner Region of the \epseri System}

In Section 5, we showed that $a_{min}\sim$1 \um gives a much better
fit to the mid-infrared disk profiles compared to the models using a
smaller size cutoff in the particle size distribution. Observationally
a wide range of $a_{min}$ has been inferred from resolved imaging and
mid-infrared spectroscopy. Since silicate-like small grains have
solid-state features in the 8--25 \um region, a featureless emission
spectrum is usually interpreted to indicate a lack of small
grains. However, the definition of ``small'' depends on the dust
composition. For example, for amorphous silicates (like astronomical
silicates), the 10 \um and 20 \um features have very similar shapes
for grain sizes of 0.05--1 \um (with a slightly sharper 10 \um feature
toward smaller sizes). As a result, at the same dust temperature the
resultant emission spectrum is very similar, and the only difference
lies in the amount of emission. Therefore, the IRS spectrum in the
\epseri inner region provides no constraint on the minimum grain size,
as also has been suggested by \citet{backman09} (who found that
the spectrum only requires $a \lesssim$3 \mm).

As discussed in \citet{reidemeister11}, the nominal blow out
size does not exist in \epseri due to its low mass and
luminosity. However, a possible blow out size might exist if pressure 
exerted by stellar winds is invoked. In a case where the contribution of
the stellar wind is strong enough, a sum of the radiation
pressure force and stellar wind pressure force could reinforce
$a_{bl}$, as has been suggested in the AU Mic disk (Figure 1 in
\citet{schuppler15}). However, for a K-star like $\epsilon$ Eri,
the effect is too small, resulting again in no blow out limit.

Theoretically, we expect some depletion in the inner region around
$\epsilon$ Eri due to enhanced drag forces (P-R and stellar wind
drags). The P-R time scale depends on the mass of the star, the
location of the dust ($R$) and the ratio between radiation force and
gravity ($\beta \propto a^{-1}$), is given as
\begin{equation} \tau_{PR} = 400\ yr \ \frac{M_{\sun}}{M_{\ast}}
(\frac{R}{AU})^2 \frac{1}{\beta}\ .
\end{equation} 
Therefore, $\tau_{PR} \sim$4 $\times10^{3}$ yr for a belt at 2 au
around $\epsilon$ Eri (assuming $\beta$ = 0.5).  Such a dust belt with
a $f_d$ of 3$\times10^{-5}$ 
has a collisional time scale of $\sim$4$\times10^{4}$ yr, roughly 10
times longer than the P-R time scale. With the additional aid of
stellar wind drag from the active star, the drag timescales will be
even shorter. Since smaller particles are dragged in faster than
larger ones, we expect a flatter size distribution at small sizes,
setting an effective $a_{min}$ larger than the typical size in a
collision-dominated system.

One can estimate $a_{min}$, or the dominant/critical grain size
($a_c$), by balancing the two source and sink time scales as suggested
by \citet{kuchner10} and \citet{wyatt11}. In the inner
region of $\epsilon$ Eri, the source time scale is the collisional
time scale, and the sink time scale is the transported time scale by
stellar wind since the stellar wind drag dominates
the P-R drag. We then re-derived Eqn.(6) of \citet{kuchner10}
as follows:
\begin{eqnarray}
a_c = 1000 \ \mu \rm{m}\ Q_{sw} \left({{\rho_g} \over {1 \  \rm{g}\ \rm{cm}^{-3}}}\right)^{-1} \left(\dot{M}_{sw} \over 30 \dot{M}_{\odot} \right) \nonumber \\
\left({M_{\ast} \over {M_{\odot}}}\right)^{-1/2}
\left({R \over {1 \ au}}\right)^{-1/2}
\left({f_d\over {10^{-7}}}\right)^{-1} ,
\end{eqnarray}
where $Q_{sw}$ is the stellar wind pressure efficiency, and $\dot{M}_{sw}$ is the stellar wind mass loss rate. 
Assuming $Q_{sw}$=1, $\rho_g$ = 2.5 g\ cm$^{-3}$, a mass loss rate of 30
times the solar value, M$_{\ast}$ = 0.82 M$_{\sun}$, $R$ = 2 au, and
$f_d$ of 3$\times 10^{-5}$, the critical grain size is
about 1 \mm.

Using the energy-conservation criterion, \citet{krijt14} analytically
derived the lower boundary of the particle sizes produced in debris
disks. Under plausible parameters, they found that $a_{min}$ could be
much larger than $a_{bl}$ depending on the collision velocity,
material parameters, and the size of the largest
fragment. \citet{thebault16} numerically investigated such an effect,
and concluded that the surface energy constraint generally has a weak
effect for early-type stars and wide (50--100 au) debris disks, but
might be more pronounced for sun-like stars and narrow belts. The
depletion of the small grains is hard to estimate since we lack
detailed information on the planetesimal belt(s) in the inner region
of $\epsilon$ Eri. As an optimal case (since some parameters of their
model are quite uncertain), the ratio between $a_{min}$ and $a_{bl}$
is $\sim$3 for a 2 au belt around \epseri (0.34 L$_{\sun}$ and 0.82
M$_{\sun}$) (Eqn.\ (7) in \citealt{krijt14}). Our choice of
$a_{min}\sim$1 \um is consistent with this limit and the critical size
estimated in the previous paragraph.

\subsection{Putative \epseri b and the Location of an Asteroid-belt Analog}

The existence of \epseri b and its orbital parameters have been
heavily debated since it was reported in 2000
\citep{hatzes00,benedict06,butler06}. The Exoplanet Encyclopedia lists
the planet at 3.5 au radius as confirmed, but \citet{zechmeister13}
combined 15 years of radial velocity data and found that it did not
support this status for the case of a highly eccentric orbit
($e\sim$0.6). A similar result was also been found by
\citet{anglada12}. \citet{howard16} analyzed all available radial
velocity measurements from Lick and Keck Observatories by the
California Planet Survey, and found that the stellar CaII H\& K
emission does not correlate with the radial velocity. Thus the radial
velocity modulation is likely caused by an external source, i.e., the
planet. Combining available ground-based high-contrast imaging,
\citet{mizuki16} presented an updated contrast curve around $\epsilon$
Eri, and could marginally rule out the 1.6 M$_J$, $e=$0.7 case
\citep{benedict06} if the age of the system is as young as 200 Myr.

As noted by \citet{backman09}, the possibility that $\epsilon$ Eri b
is on a highly eccentric orbit is inconsistent with the existence of
an inner warm debris belt at $\sim$3 au. If the planet co-exists with
an asteroid-belt analog (i.e., one within 2 au), its orbit must have
low eccentricity ($e \le$0.2), based on the dynamical stability study
by \citet{brogi09}. We estimated the width of the planet's chaotic
zone by assuming $\epsilon$ Eri b is 1.6--1.7 M$_J $ with
$e$= 0--0.2. The inner boundary of the chaotic zone is then at 2--2.6
au, with 3.5--5 au for the outer boundary using the formulae from
\citet{mustill12} and \citet{morrison15}. Any planetesimal belt in the
inner region of the $\epsilon$ Eri system must be located inside 2 au
and/or outside 5 au to be dynamically stable with the assumed
$\epsilon$ Eri b. For this reason, we constructed the asteroid-belt
analog at 1.5--2 au in Section 5.3.

If there is no $\epsilon$ Eri b, the location of the asteroid-belt
could be at 3 au as proposed by \citet{backman09} using the dust
temperature argument. Since current data put no constraints on the
exact number of planetesimal belts (one or two) in the inner
$\epsilon$ Eri region, the inner warm component could be a dragged-in
component from the outer {\it warm} planetesimal belt ($\sim$8--20
au). This remains a possibility because the P-R time scale is similar
to the collisional time scale in a 10-au belt around $\epsilon$ Eri
(and would be shorter with the aid of stellar wind drag).

\subsection{Expected Millimeter Emission from the Inner \epseri Region}

Although the new {\it SOFIA} data rule out the drag-dominated
transported model for the inner debris, another version of the
transported dust scenario, involving disintegration of icy comets
scattered inward by the planets interior of a cold planetesimal region
\citep{morales11,bonsor12}, remains plausible for the source of inner
debris. As the perturbed icy planetesimals get closer to the ice line,
sublimation of volatile material likely causes them to become
active comets, populating the inner region with dust.  This mechanism
is found to be the primary source of the warm dust in the inner region
of the solar system \citep{nesvorny10,ueda17}. However, there is
circumstantial evidence that the warm excesses in some other systems
are aligned with their primordial ice lines, not the current-day ones,
and thus the dust arises from in-situ planetesimal
belts (Ballering et al.\ 2017, submitted). Comets release material
primarily in the form of coarse mm- to cm-sized dust grains, and the
radial span of the big grains from active comets should be broad, as
is the radial distribution of comets themselves. For an in-situ
planetesimal belt, the distribution of large dust grains should be
narrower, because the eccentricities of the parent bodies are expected
to be lower than those of comets. As a result, the mm emission from
the in-situ planetesimal belt should be confined in a narrower
distribution, and thus be more easily detectable, than the one from
the cometary grains.  Therefore, the ultimate test to differentiate
the in-situ and transported origins of the warm dust is to detect the
submm/mm emission from large grains in the inner region.

\epseri has been observed by many radio single-dish and
interferometric facilities, which provide some constraints on the
amount of submm/mm emission in the inner region. \epseri is a young
and active star, and is known to possess excess free-free emission at
7 mm \citep{macgregor15}, making it difficult to evaluate the
possible excess emission at mm wavelengths using integrated
photometry. No confirmed excess emission is reported by
\citet{lestrade15} and \citet{macgregor15} at 1.2/1.3 mm, while
\citet{chavez16} report an ($\sim$5 $\sigma$) excess of 1.3 mJy at 1.1
mm in the inner 18 au region. However, the high resolution ALMA 1.3 mm
image of the system rules out any narrow belt with a total flux
density of $>$0.8 mJy in the inner region (Booth et al.\ 2017,
submitted). The ALMA observation was one single pointing offset from
the star designed to image the northern part of the cold ring,
providing poor coverage of the inner region. The mm detection of the
inner debris remains controversial.

We predict the flux density of the inner debris at 1.3 mm using the
three tested models in Section 5. For the dragged-in model, the mm
flux is much less than 10 $\mu$Jy due to lack of large grains. For the
puffed-up disk model and the inner 1.5--2 au belt model (asteroid-belt
analog), the total flux density is $\lesssim$16 $\mu$Jy at 1.3 mm,
making the mm detection very challenging.  The outer warm belt model
(8--20 AU) gives a total flux density of 0.7 mJy at 1.3 mm, comparable
to the expected stellar emission. Given the difficulty of predicting
the stellar output in the mm wavelengths, the best way to confirm
such a belt is to resolve it from the star. If the outer warm belt is
narrow in the mm wavelengths, a belt at 13 au ($\sim$4\arcsec\ radius)
would have a surface brightness of 27 $\mu$Jy/beam at 1.3 mm assuming
a beam of 1\arcsec\ and an unresolved width. The ALMA
observation obtained by Booth et al.\ (submitted) reaches a rms of 14
$\mu$Jy/beam. Under nominal conditions (e.g., if the source region was
centered at the primary beam), the proposed outer warm belt could be
detected at $\sim$2 $\sigma$, making a confirmed detection ($>$3 $\sigma$)
difficult. If the outer warm belt is slightly larger or broader than the
beam size, the expected surface brightness would be less as a result.

\section{Conclusion}

We obtained a {\it SOFIA}/FORCAST resolved image of $\epsilon$ Eri and
confirmed the presence of excess emission coinciding with the star at
35 $\mu$m. The excess emission is resolved by $\sim$2 beam widths
(FWHM $\sim$3\farcs4), suggesting the emission region is extended
beyond $\sim$10 au. We derived the 35 $\mu$m disk radial profile for
$\epsilon$ Eri, and found that the emission region is consistent with
either (1) a broad, centrally peaked, Gaussian profile structure with
a width (FWHM) of 18 au or (2) an unresolved central source plus a Gaussian
cross-section ring peaked at 10 au with a width of 10 au
(unresolved). To further characterize the amount and structure of
the excess in the \epseri inner region, we also re-analyzed the 
previously published {\it Spitzer} IRS and MIPS 24 \um data. These 
observations represent the best data sets that can be used to test 
the origin of the warm excess in the \epseri system. 

Using the FORCAST 35 and MIPS 24 $\mu$m disk profiles, we tested three
different dust distributions in the inner 25 au region of $\epsilon$
Eri to probe the nature of the warm excess. We found that the presence
of in-situ dust-producing planetesimal belt(s) is the most likely
source of the excess emission, and that the current data cannot distinguish
between one broad (3--21 au) puffed-up disk and two separate
planetesimal belts. In the two distinct belt case, the outer warm disk
can be as close as $\sim$8 au and extend up to 20 au. Furthermore,
the inner warm disk can be a true asteroid-belt analog (i.e., a
planetesimal belt located near the ice line) at 1.5-2 au, which is
consistent with the presence of \epseri b as long as its orbit is
nearly circular. The high resolution of the {\it SOFIA } data enables
us to differentiate the in-situ dust source from grains under the
influence of P-R and stellar wind drags. The newly obtained 35 $\mu$m
disk profile is not consistent with the drag-dominated case (constant
dust flow from the outer (64 au) cold Kuiper-belt analog); however, a
contribution from a small amount of dragged-in grains cannot be ruled
out.

\acknowledgments

Based in part on observations made with the NASA/DLR Stratospheric
Observatory for Infrared Astronomy ({\it SOFIA}). {\it SOFIA} is
jointly operated by the Universities Space Research Association,
Inc. (USRA), under NASA contract NAS2-97001, and the Deutsches SOFIA
Institut (DSI) under DLR contract 50 OK 0901 to the University of
Stuttgart. Financial support for this work was provided by NASA
through award \# SOF02-0061 and SOF03-0092 issued by USRA. KYLS
acknowledges the partial support from the NASA grant \# NNX15AI86G,
and the data reduction help from the {\it SOFIA} Science Center. AVK
and TL acknowledge support by the DFG, grants Kr 2164/13-1, Kr
2164/15-1 and Lo 1715/2-1.

\medskip 

\noindent Facilities: \facility{SOFIA(FORCAST), Spitzer(MIPS, IRS)}.


\begin{thebibliography}{}


\bibitem[Anglada-Escud{\'e} \& Butler(2012)]{anglada12} Anglada-Escud{\'e}, G., \& Butler, R.~P.\ 2012, \apjs, 200, 15 

\bibitem[Backman et al.(2009)]{backman09} Backman, D., Marengo, M., Stapelfeldt, K., et al.\ 2009, \apj, 690, 1522 


\bibitem[Ballering et al.(2016)]{ballering16} Ballering, N.~P., Su, K.~Y.~L., Rieke, G.~H., \& G{\'a}sp{\'a}r, A.\ 2016, \apj, 823, 108 

\bibitem[Benedict et al.(2006)]{benedict06} Benedict, G.~F., McArthur, B.~E., Gatewood, G., et al.\ 2006, \aj, 132, 2206 

\bibitem[Bonsor \& Wyatt(2012)]{bonsor12} Bonsor, A., \& Wyatt, M.~C.\ 2012, \mnras, 420, 2990 


\bibitem[Brogi et al.(2009)]{brogi09} Brogi, M., Marzari, F., \& Paolicchi, P.\ 2009, \aap, 499, L13 

\bibitem[Burns et al.(1979)]{burns79} Burns, J.~A., Lamy, P.~L., \& Soter, S.\ 1979, {\it Icarus}, 40, 1 

\bibitem[Butler et al.(2006)]{butler06} Butler, R.~P., Wright, J.~T., Marcy, G.~W., et al.\ 2006, \apj, 646, 505 

\bibitem[Chavez-Dagostino et al.(2016)]{chavez16} Chavez-Dagostino, M., Bertone, E., Cruz-Saenz de Miera, F., et al.\ 2016, \mnras, 462, 2285 

\bibitem[Deller \& Maddison(2005)]{deller05} Deller, A.~T., \& Maddison, S.~T.\ 2005, \apj, 625, 398 


\bibitem[Di Folco et al.(2004)]{difolco04} Di Folco, E., Th{\'e}venin, F., Kervella, P., et al.\ 2004, \aap, 426, 601 


\bibitem[Engelbracht et al.(2007)]{engelbracht07} Engelbracht, C.~W., Blaylock, M., Su, K.~Y.~L., et al.\ 2007, \pasp, 119, 994 

\bibitem[G{\'a}sp{\'a}r et al.(2012)]{gaspar12} G{\'a}sp{\'a}r, A., Psaltis, D., Rieke, G.~H., \& {\"O}zel, F.\ 2012, \apj, 754, 74 

\bibitem[Gray et al.(2003)]{gray03} Gray, R.~O., Corbally, C.~J., Garrison, R.~F., McFadden, M.~T., \& Robinson, P.~E.\ 2003, \aj, 126, 2048 

\bibitem[Gray et al.(2006)]{gray06} Gray, R.~O., Corbally, C.~J., Garrison, R.~F., et al.\ 2006, \aj, 132, 161 

\bibitem[Gehrz et al.(2009)]{gehrz09} Gehrz, R.~D., Becklin, E.~E., de Pater, I., et al.\ 2009, Advances in Space Research, 44, 413 

\bibitem[Gordon et al.(2005)]{gordon05} Gordon, K.~D., Rieke, G.~H., Engelbracht, C.~W., et al.\ 2005, \pasp, 117, 503 

\bibitem[Greaves et al.(1998)]{greaves98} Greaves, J.~S., Holland, W.~S., Moriarty-Schieven, G., et al.\ 1998, \apjl, 506, L133 

\bibitem[Greaves et al.(2014)]{greaves14} Greaves, J.~S., Sibthorpe, B., Acke, B., et al.\ 2014, \apjl, 791, L11 

\bibitem[Hatzes et al.(2000)]{hatzes00} Hatzes, A.~P., Cochran, W.~D., McArthur, B., et al.\ 2000, \apjl, 544, L145 


\bibitem[Herter et al.(2012)]{herter12} Herter, T.~L., Adams, J.~D., De Buizer, J.~M., et al.\ 2012, \apjl, 749, L18 

\bibitem[Howard \& Fulton(2016)]{howard16} Howard, A.~W., \& Fulton, B.~J.\ 2016, \pasp, 128, 114401 


\bibitem[Janson et al.(2015)]{janson15} Janson, M., Quanz, S.~P., Carson, J.~C., et al.\ 2015, \aap, 574, A120 

\bibitem[Kennedy \& Kenyon(2008)]{kennedy08} Kennedy, G.~M., \& Kenyon, S.~J.\ 2008, \apj, 673, 502-512 

\bibitem[Kennedy \& Wyatt(2014)]{kennedy14} Kennedy, G.~M., \& Wyatt, M.~C.\ 2014, \mnras, 444, 3164 

\bibitem[Krijt \& Kama(2014)]{krijt14} Krijt, S., \& Kama, M.\ 2014, \aap, 566, L2 

\bibitem[Kuchner \& Stark(2010)]{kuchner10} Kuchner, M.~J., \& Stark, C.~C.\ 2010, \aj, 140, 1007 

\bibitem[Laor \& Draine(1993)]{laor93} Laor, A., \& Draine, B.~T. 1993, \apj, 402, 441

\bibitem[Lebouteiller et al.(2015)]{lebouteiller15} Lebouteiller, V., Barry, D.~J., Goes, C., et al.\ 2015, \apjs, 218, 21 


\bibitem[Lestrade \& Thilliez(2015)]{lestrade15} Lestrade, J.-F., \& Thilliez, E.\ 2015, \aap, 576, A72 

\bibitem[MacGregor et al.(2015)]{macgregor15} MacGregor, M.~A., Wilner, D.~J., Andrews, S.~M., Lestrade, J.-F., \& Maddison, S.\ 2015, \apj, 809, 47 

\bibitem[Mamajek \& Hillenbrand(2008)]{mamajek08} Mamajek, E.~E., \& Hillenbrand, L.~A.\ 2008, \apj, 687, 1264-1293 

\bibitem[Matthews et al.(2014)]{matthews14} Matthews, B.~C., Krivov, A.~V., Wyatt, M.~C., Bryden, G., \& Eiroa, C.\ 2014, Protostars and Planets VI, 521 

\bibitem[Mizuki et al.(2016)]{mizuki16} Mizuki, T., Yamada, T., Carson, J.~C., et al.\ 2016, \aap, 595, A79 

\bibitem[Morales et al.(2011)]{morales11} Morales, F.~Y., Rieke, G.~H., Werner, M.~W., et al.\ 2011, \apjl, 730, L29 

\bibitem[Morrison \& Malhotra(2015)]{morrison15} Morrison, S., \& Malhotra, R.\ 2015, \apj, 799, 41 

\bibitem[Mustill \& Wyatt(2012)]{mustill12} Mustill, A.~J., \& Wyatt, M.~C.\ 2012, \mnras, 419, 3074 
\bibitem[Nesvorn{\'y} et al.(2010)]{nesvorny10} Nesvorn{\'y}, D., Jenniskens, P., Levison, H.~F., et al.\ 2010, \apj, 713, 816 

\bibitem[Ozernoy et al.(2000)]{ozernoy00} Ozernoy, L.~M., Gorkavyi, N.~N., Mather, J.~C., \& Taidakova, T.~A.\ 2000, \apjl, 537, L147 

\bibitem[Quillen(2006)]{quillen06} Quillen, A.~C.\ 2006, \mnras, 372, L14

\bibitem[Quillen \& Thorndike(2002)]{quillen02} Quillen, A.~C., \& Thorndike, S.\ 2002, \apjl, 578, L149 

\bibitem[Reidemeister et al.(2011)]{reidemeister11} Reidemeister, M., Krivov, A.~V., Stark, C.~C., et al.\ 2011, \aap, 527, A57 

\bibitem[Rieke et al.(2008)]{rieke08} Rieke, G. H.; Blaylock, M.; Decin, L.; Engelbracht, C. et al. 2008, AJ, 135, 2245

\bibitem[Sch{\"u}ppler et al.(2015)]{schuppler15} Sch{\"u}ppler, C., L{\"o}hne, T., Krivov, A.~V., et al.\ 2015, \aap, 581, A97 

\bibitem[Sierchio et al.(2014)]{sierchio14} Sierchio, J.~M., Rieke, G.~H., Su, K.~Y.~L., \& G{\'a}sp{\'a}r, A.\ 2014, \apj, 785, 33 


\bibitem[Stark(2011)]{stark11} Stark, C.~C.\ 2011, \aj, 142, 123 

\bibitem[Su et al.(2013)]{su13} Su, K.~Y.~L., Rieke, G.~H., Malhotra, R., et al.\ 2013, \apj, 763, 118 

\bibitem[Su et al.(2016)]{su16} Su, K.~Y.~L., Rieke, G.~H., Defr{\'e}re, D., et al.\ 2016, \apj, 818, 45 

\bibitem[Thebault(2016)]{thebault16} Thebault, P.\ 2016, \aap, 587, A88 


\bibitem[Ueda et al.(2017)]{ueda17} Ueda, T., Kobayashi, H., Takeuchi, T., et al.\ 2017, arXiv:1702.03086 


\bibitem[van Leeuwen(2007)]{vanleeuwen07} van Leeuwen, F.\ 2007, \aap, 474, 653 

\bibitem[Wood et al.(2002)]{wood02} Wood, B.~E., M{\"u}ller, H.-R., Zank, G.~P., \& Linsky, J.~L.\ 2002, \apj, 574, 412 

\bibitem[Wyatt(2006)]{wyatt06} Wyatt, M.~C.\ 2006, \apj, 639, 1153 

\bibitem[Wyatt et al.(2007)]{wyatt07} Wyatt, M.~C., Smith, R.,
Greaves, J.~S., et al.\ 2007, \apj, 658, 569

\bibitem[Wyatt et al.(2011)]{wyatt11} Wyatt, M.~C., Clarke, C.~J., \& Booth, M.\ 2011, Celestial Mechanics and Dynamical Astronomy, 111, 1 


\bibitem[Young et al.(2012)]{young12} Young, E.~T., Becklin, E.~E., Marcum, P.~M., et al.\ 2012, \apjl, 749, L17 

\bibitem[Zechmeister et al.(2013)]{zechmeister13} Zechmeister, M., K{\"u}rster, M., Endl, M., et al.\ 2013, \aap, 552, A78 



\end{thebibliography}
\end{document}